Preprint

# The spike-timing-dependent learning rule to encode spatiotemporal patterns in a network of spiking neurons

Masahiko Yoshioka*

*Brain Science Institute, RIKEN,*

*Hirosawa 2-1, Wako-shi, Saitama, 351-0198, Japan*

**Abstract**

We study associative memory neural networks based on the Hodgkin-Huxley type of spiking neurons. We introduce the spike-timing-dependent learning rule, in which the time window with the negative part as well as the positive part is used to describe the biologically plausible synaptic plasticity. The learning rule is applied to encode a number of periodical spatiotemporal patterns, which are successfully reproduced in the periodical firing pattern of spiking neurons in the process of memory retrieval. The global inhibition is incorporated into the model so as to induce the gamma oscillation. The occurrence of gamma oscillation turns out to give appropriate spike timings for memory retrieval of discrete type of spatiotemporal pattern. The theoretical analysis to elucidate the stationary properties of perfect retrieval state is conducted in the limit of an infinite number of neurons and shows the good agreement with the result of numerical simulations. The result of this analysis indicates that the presence of the negative and positive parts in the form of the time window contributes to reduce the size of crosstalk term, implying that the time window with the negative and positive parts is suitable to encode a number of spatiotemporal patterns. We draw some phase diagrams, in which we find various types of phase transitions with change of the intensity of global inhibition.

## 1 Introduction

In the past few decades there has been some theoretical interest in associative memory neural networks [1–4]. A major breakthrough was made by Hopfield, who has introduced the stochastic neural network model with an energy function [5]. By means of the method based on the statistical mechanical theory several authors have conducted the investigations on Ising spin networks [6–12] and analog neural networks [13–18], which have clarified much of the fundamental properties of associative memory neural networks.

Meanwhile, in electrophysiological experiments, a significant effort has been devoted to clarify the capability of the real nervous system to memorize spatiotemporal patterns [19, 20]. Recently, it has been revealed that in the long spike sequences of the rat hippocampus short spike sequences appear repeatedly [21]. This phenomenon imply the capability of the rat hippocampus to memorize spatiotemporal patterns on the basis of spike timings, and hence, concern has been raised about associative memory neural network models in which information is represented by spike timings of neurons [22, 23].

To deal with the problem concerning spike timings of neurons one might consider investigating networks of simple phase oscillators. Since some theoretical analysis is available, the stationary properties of associative memory based on networks of simple oscillators have been studied extensively both in the case of an extensive number of stored patterns and in the case of distributed natural frequencies [24–28]. Even in the presence of white noise as well as a distribution of natural frequencies we can derive the storage capacity of networks of phase oscillators analytically [29].

*e-mail: myosioka@brain.riken.go.jp

For a more complete understanding of the information processing based on spike timings of neurons, it is, however, necessary to adopt more biologically plausible neural network models because such features as the time evolution of membrane potentials and decay time of synaptic electric currents play a significant role in the rhythmic behavior of neurons. For this purpose, networks of spiking neurons are considered to be suitable models for investigation, though it remains an unsolved problem to find the adequate learning rule to encode spike timings in networks of spiking neurons.

Since networks of spiking neurons with asymmetric synaptic connections exhibit sequential firings of neurons [30,31], one may consider that the learning rule to encode spatiotemporal patterns should generate asymmetric synaptic connections. Actually, incorporating asymmetric synaptic connections, Gerstner et al. has investigated the networks of the integrated-and-fire type of spiking neurons with discrete time dynamics, in which the encoded spatiotemporal patterns are successfully reproduced in spike timings of neurons in the process of memory retrieval [32]. Then, the question arises as to how such asymmetric synaptic connections are developed in a real nervous system. The results of the recent electrophysiological experiments have revealed that the modification of a excitatory synaptic efficacy depends on the precise timings of pre- and postsynaptic firings [33–35]. A synaptic efficacy is found to increase if firing of a presynaptic neuron occurs in advance of firing of a postsynaptic neuron, and to decrease otherwise. Accordingly, the time window to describe the spike-timing-dependent synaptic plasticity takes the form having the negative and positive parts as is described in Fig. 1. Several authors have proposed that this modification rule serves to solve such the problems as path navigation [36,37], direction selectivity [38,39], competitive Hebbian learning [40], and biologically plausible derivation of the Linsker's equation as well as the Hebbian learning rule [41]. In the present study, we aim to tackle the problem of how spatiotemporal patterns are encoded in a network of spiking neurons on the basis of the spike-timing-dependent modification rule. We introduce the spike-timing-dependent learning rule, which gives asymmetric synaptic connections so that networks of spiking neurons function as associative memory.

Spiking neurons we assume in the present study interact with each other without time delay, that is, every neuron obtains synaptic electric current immediately after one neuron fires. In this case, the sequential firings of neurons for memory retrieval take place with rather short time intervals, and one might consider such rapid pattern retrieval makes no sense from a biological point of view. It may be desirable that the network equips a certain mechanism to control spike timings of neurons to realize the information processing with the adequate processing period.

We hypothesize that the gamma oscillation is the key mechanism to solving this problem. In the various regions of real nervous system, such as the neocortex and the hippocampus, a population of neurons are found to exhibit synchronized firings with a characteristic frequency 20-80Hz, and such synchronized firings of neurons, namely the gamma oscillation, attract much attention of researchers [42–48]. When the gamma oscillation arises, firings of neurons occur only around discrete time steps, and the situation is somewhat similar with the case of the Hopfield model with the discrete time dynamics. We hypothesize that such the discrete type of firing pattern serves to control spike timings of neurons. Some experimental and theoretical results support the hypothesis that the global inhibition, which is induced by the presence of interneurons, plays a significant role in generation of the gamma oscillation [49–55]. In the present study, incorporating the global inhibition into the model, we aim to investigate the influence of the gamma oscillations on the properties of memory retrieval.

It should be noted that we can apply some theoretical techniques to analyze the stationary properties of the present system provided that the number of encoded patterns are sufficiently small (i.e. $P/N \ll 1$, where $P$ is the number of encoded patterns and $N$ is the number of neurons). When retrieval is successful, the periodical behavior of every neuron is identical, but shifts with respect to time depending on the value of the target pattern, and thus we can reduce the many body problem into the single body problem in the limit of an infinite number of neurons. By use of this exact reduction we can draw some phase diagrams, which clarify the condition for successful retrieval and the occurrence of phase transitions. Furthermore, this method of analysis leads us to find one surprising property of the present system: the crosstalk term vanishes if the area of the positive part of the time window is equivalent to the area of the negative part so that the time integration of the time window takes value zero. This result implies that the present form of the time window, which has the negative and positive parts, is suitable to encode a number of spatiotemporal patterns.

The present paper is organized as follows. In section 2, we present the details of the neural network model we study, and then we introduce the spike-timing-dependent learning rule to encode spatiotemporal patterns. In section 3, we investigate the stationary properties of the network in perfect retrieval state analytically. In the course of this analysis, it becomes clear that the negative and positive parts of the time window play an important role in reducing the size of crosstalk term. In section 4, we apply this method of analysis to the case with continuous type of patterns to clarify the condition for the occurrence of the perfect retrieval. The result of the numerical simulations are presented showing good agreement with the result of the theoretical analysis. Then, in section 5, we treat the case of discrete type of patterns, which are successfully retrieved when the gamma oscillation arises. Finally, in section 6, we give a brief summary of the present study.

## 2 Model of a network of spiking neurons

In real nervous system some regions, such as the neocortex and the hippocampus, are found to comprise a large number of pyramidal cells and interneurons. In these networks pyramidal cells typically connect to other neurons (i.e., both pyramidal cells and interneurons) via excitatory synapses, while interneurons connect to pyramidal cells via GABAergic synapses (inhibitory synapses). When one pyramidal cell fires, the other pyramidal cells obtain excitatory postsynaptic potential (EPSP) due to the excitatory synapses that connect pyramidal cell to the other pyramidal cells. At the same time, some interneurons surrounding the firing pyramidal cell also obtain EPSP due to the excitatory synapses that connect the pyramidal cell to interneurons. Since the threshold value for firing of interneurons is rather small, these interneurons begin to fire immediately after the arrival of action potentials from the firing pyramidal cell, and then such firings of the interneurons give rise to the inhibitory postsynaptic potentials (IPSPs) into a large number of pyramidal cells via GABAergic synapses. In this way, when one pyramidal cell fires, the other pyramidal cells obtain two kinds of post synaptic potentials: EPSP induced by the direct arrival of action potential from the firing pyramidal cell and IPSPs mediated by firings of interneurons surrounding the firing pyramidal cell.

For the purpose of elucidating the fundamental properties of the nervous system composed of pyramidal cells and interneurons, we investigate a network of $N$ spiking neurons interacting through two types of synaptic electric currents, namely, electric currents via plastic synapses $J_{ij}$ and global inhibition. The dynamics of a network of spiking neurons we study is expressed in the form

$$\dot{V}_i = f\left(V_i, W_{i1}, \ldots, W_{in}\right) + I_i(t), \tag{1}$$

$$\dot{W}_{ij} = g_j\left(V_i, W_{i1}, \ldots, W_{in}\right), \quad i = 1, \ldots, N, \; j = 1, \ldots, n \tag{2}$$

with

$$I_i(t) = I_{pyr,i}(t) + I_{int}(t) + I_{ext,i}(t), \tag{3}$$

where $V_i(t)$ denotes the membrane potential of neuron $i$ and $W_{ij}(t)$ auxiliary variables necessary for neurons to exhibit spiking behavior. The definition of the electric currents $I_{pyr,i}(t)$, $I_{int}(t)$, and $I_{ext,i}(t)$ will be explained in what follows. Note that now we focus on the dynamics of a network of $N$ pyramidal cells and omit describing the detailed dynamics of interneurons [56]. For the dynamics $f\left(V, W_1, \ldots, W_n\right)$ and $g_j\left(V, W_1, \ldots, W_n\right)$, several authors have assumed the Hodgkin-Huxley equations [57], the FitzHugh-Nagumo equations [58, 59], and so on. In the present study we assume the Hodgkin-Huxley equations, and hence the degrees of freedom of a state of a neuron is 4 (i.e., $n = 3$). In appendix A, we present the details of the Hodgkin-Huxley equations we adopt in the present study.

$I_{pyr,i}(t)$ denotes a sum of synaptic electric currents via plastic synapses $J_{ij}$, which is activated by the arrival of action potential from other pyramidal cells. We define firing times of neuron $i$ as the time when the membrane potential $V_i(t)$ exceeds the threshold value $V_0 = 0$ and denote $k$-th firing time of neuron $i$ by $t_i(k)$. Then, the synaptic electric current $I_{pyr,i}(t)$ is written in the form

$$I_{pyr,i}(t) = A_{pyr} \sum_{j=1}^{N} \sum_{k} J_{ij} S_{pyr}\left\{t - t_j(k)\right\}, \qquad i = 1, \ldots, N \tag{4}$$

where $J_{ij}$ denotes a synaptic efficacy from neuron $j$ to neuron $i$, and $A_{pyr}$ is the variable controlling the intensity of synaptic electric current $I_{pyr,i}(t)$. We assume the time-dependent postsynaptic potential $S_{pyr}(t)$ of the form

$$S_{pyr}(t) = \begin{cases} 0 & t < 0 \\ \dfrac{1}{\tau_{pyr,1} - \tau_{pyr,2}} \left\{ \exp\left(-\dfrac{t}{\tau_{pyr,1}}\right) - \exp\left(-\dfrac{t}{\tau_{pyr,2}}\right) \right\} & 0 \leq t \end{cases}. \tag{5}$$

In what follows, we set $\tau_{pyr,1} = 10$ (msec) and $\tau_{pyr,2} = 5$ (msec).

For the sake of brevity, instead of describing the detailed dynamics of interneurons, we simply assume that IPSPs are induced in other pyramidal cells immediately after one of $N$ pyramidal cells fires [56], that is, we assume global inhibition $I_{int}(t)$ of the form

$$I_{int}(t) = \frac{A_{int}}{N} \sum_{j=1}^{N} \sum_{k} S_{int} \left\{ t - t_j(k) \right\}, \tag{6}$$

where $A_{int}$ is the variable controlling the intensity of global inhibition. Note that the global inhibition $I_{int}(t)$ is independent of index $i$ since every neuron obtains the same amount of inhibitory electric current. The time-dependent inhibitory postsynaptic potential $S_{int}(t)$ is described as

$$S_{int}(t) = \begin{cases} 0 & t < 0 \\ \dfrac{-1}{\tau_{int,1} - \tau_{int,2}} \left\{ \exp\left(-\dfrac{t}{\tau_{int,1}}\right) - \exp\left(-\dfrac{t}{\tau_{int,2}}\right) \right\} & 0 \leq t \end{cases}. \tag{7}$$

Note that $S_{int}(t)$ takes negative value in the interval $0 \leq t$. In what follows, we set $\tau_{int,1} = 5$ (msec) and $\tau_{int,2} = 2.5$ (msec) so that $S_{int}(t)$ decays faster than $S_{pyr}(t)$.

The external electric current $I_{ext,i}(t)$ is used to control initial firings of neurons. For the initial condition of the network, we set state of neurons $(V_i, \{W_i\})$ $(i = 1, \dots, N)$ to be at the stable fixed point of the dynamics of Eqs. (1) and (2) with $I_i(t) = 0$. It means that, without any external stimuli, all neurons keep quiescent irrespective of the strength of synaptic efficacy $J_{ij}$. Thus, for the purpose of invoking initial firings that act as a trigger to retrieve the target pattern, we use the pulsed form of the external electric current $I_{ext,i}(t)$ only at the beginning of the evolution of the dynamics. Then, the initial firings of neurons invoke the synaptic electric currents, which become driving force for the next firings of neurons. Once note that the external electric current $I_{ext,i}(t)$ is used only at the beginning. In the theoretical analysis below we set $I_{ext,i}(t) = 0$ because we focus on the stationary behavior in this analysis.

The aim of considering the present model is to investigate the properties of nervous system composed of pyramidal cells and interneurons. As is mentioned above, in real nervous system, pyramidal cells are found to interact with other pyramidal cells via excitatory synapses. Nevertheless, in what follows, for the purpose of simplifying the situation, we assume synaptic efficacy $J_{ij}$ can take not only positive value but also negative value. This assumption might be somewhat implausible, but allow one to introduce the simple learning rule, which is amenable to satisfactory level of analysis. In the next subsection, on the basis of naive assumption that this simplification does not change the fundamental properties of the system, we introduce the learning rule that is capable of encoding a number of spatiotemporal patterns.

## 2.1 The spike-timing-dependent learning rule to encode spatiotemporal patterns

The periodical spatiotemporal patterns we study in the present study are generated randomly with the constraint that every neuron fires only once in one period. We represent these spatiotemporal patterns by use of the firing times $s_i^\mu \in [0, T)$ $(i = 1, \dots, N,\ \mu = 1, \dots, P)$, where $P$ is a number of patterns and $T$ is a period of spatiotemporal patterns. To choose $s_i^\mu$ randomly from the interval $[0, T)$ we use the equation:

$$s_i^\mu = \frac{T}{Q} q_i^\mu, \qquad i = 1, \dots, N,\ \mu = 1, \dots, P, \tag{8}$$

where $Q$ is a natural number controlling the degree of discreteness of spatiotemporal patterns and random integer $q_i^\mu$ is chosen from the interval $[0, Q)$ with equal probability. We term random patterns with finite $Q$ discrete type of patterns, while we term those with $Q \to \infty$ continuous type of patterns. In what follows, setting $T = 100$, we study the case of discrete type of patterns ($Q = 10$) as well as continuous type patterns ($Q \to \infty$). By considering the case of discrete type of patterns we aim to study the effect of the occurrence of the gamma oscillation in the learning process. For convenience of the calculation below, we introduce the phase variables $\theta_i^\mu$ defined as

$$\theta_i^\mu = \frac{2\pi}{Q} q_i^\mu = \frac{2\pi}{T} s_i^\mu. \tag{9}$$

To find a clue to encode spatiotemporal patterns in a network of spiking neurons, we begin with estimating the modification of synaptic efficacy assuming that neurons fire periodically according to one of the spatiotemporal patterns. The results of the recent electrophysiological experiments suggest that the modification of a synaptic efficacy depends on the precise timings of presynaptic and postsynaptic firings [33–35], and such modification of synaptic efficacy $\Delta J$ is approximately written in the form

$$\Delta J \propto W(\Delta t) = \begin{cases} \dfrac{-1}{\tau_{W,1} - \tau_{W,2}} \left\{ \exp\left(\dfrac{\Delta t}{\tau_{W,1}}\right) - \exp\left(\dfrac{\Delta t}{\tau_{W,2}}\right) \right\} & \Delta t < 0 \\ \dfrac{1}{\tau_{W,1} - \tau_{W,2}} \left\{ \exp\left(-\dfrac{\Delta t}{\tau_{W,1}}\right) - \exp\left(-\dfrac{\Delta t}{\tau_{W,2}}\right) \right\} & 0 \le \Delta t \end{cases} \tag{10}$$

with

$$\Delta t = t_{post} - t_{pre}, \tag{11}$$

where $t_{post}$ and $t_{pre}$ denote firing times of presynaptic and postsynaptic neurons respectively. In what follows, we set $\tau_{W,1} = 10$ (msec) and $\tau_{W,2} = 5$ (msec), with which the time window $W(\Delta t)$ takes the form described in Fig. 1. When neurons fire periodically according to pattern 1, namely $s_i^1$, the firing times of neurons are written in the form

$$\tilde{s}_i^1(k) = s_i^1 + kT, \qquad i = 1, \ldots, N, \; k = \ldots, -2, -1, 0, 1, 2, \ldots \tag{12}$$

Substituting Eq. (12) into Eq. (10) we obtain the rough estimation of the modification of synaptic efficacy:

$$\begin{aligned} &\Delta J_{ij} \\ \propto\; & \sum_{k_i=-\infty}^{\infty} \sum_{k_j=-\infty}^{\infty} W\left\{\tilde{s}_i^1(k_i) - \tilde{s}_j^1(k_j)\right\} \\ =\; & \sum_{k_i} \sum_{k_j} W\left(s_i^1 - s_j^1 + k_i T - k_j T\right) \\ \propto\; & \sum_{k=-\infty}^{\infty} W\left(s_i^1 - s_j^1 + kT\right) \\ =\; & \tilde{W}\left(s_i^1 - s_j^1\right), \end{aligned} \tag{13}$$

where the periodical function $\tilde{W}(\Delta t + T) = \tilde{W}(\Delta t)$ is defined as

$$\tilde{W}(\Delta t) = \sum_{k=-\infty}^{\infty} W(\Delta t + kT). \tag{14}$$

Substituting Eq. (10) into Eq. (14) yields the explicit form of the function $\tilde{W}(\Delta t)$:

$$\begin{aligned} \tilde{W}(\Delta t) \;=\; & \frac{1}{\tau_{W,1} - \tau_{W,2}} \left\{ \frac{e^{-\Delta t/\tau_{W,1}} - e^{-(T-\Delta t)/\tau_{W,1}}}{1 - e^{-T/\tau_{W,1}}} - \frac{e^{-\Delta t/\tau_{W,2}} - e^{-(T-\Delta t)/\tau_{W,2}}}{1 - e^{-T/\tau_{W,2}}} \right\}. \\ & 0 \le \Delta t < T. \end{aligned} \tag{15}$$

In the above estimation we have treated the case with only a single spatiotemporal pattern. Now we would like to extend this result to the case with a number of spatiotemporal patterns. Since the total change of synaptic efficacy is assumed to be given by the sum of the individual changes, we extend Eq. (14) to the form

$$J_{ij} = \frac{1}{N} \sum_{\mu=1}^{P} \tilde{W}\left(s_i^\mu - s_j^\mu\right) = \frac{1}{N} \sum_{\mu=1}^{P} \tilde{W}\left\{\frac{T}{2\pi}\left(\theta_i^\mu - \theta_j^\mu\right)\right\}, \tag{16}$$

where we take proper scaling with respect to $N$. This is the learning rule we adopt in the present study. In what follows, we investigate networks of spiking neurons in which synaptic efficacy $J_{ij}$ is given by Eq. (16). As will be shown in the next section, spatiotemporal patterns encoded by use of the learning rule (16) are retrieved successfully if we give an appropriate initial condition.

# 3 Analysis of the stationary properties of perfect retrieval state in the case of a finite number of encoded patterns

The present neural networks happen to show rich variety of dynamical behavior depending on the value of the parameters such as $A_{pyr}$ and $A_{int}$. Among these behavior the most important one may be pattern retrieval in which every neuron fires periodically according to one of the encoded patterns. In such a case, firing times of neurons are written in the form

$$t_i(k) = \frac{\tilde{T}}{2\pi}\theta_i^1 + k\tilde{T}, \qquad i = 1,\ldots,N,\ k = \ldots,-2,-1,0,1,2,\ldots, \tag{17}$$

where we chose pattern 1 as the retrieved one. Note that, in general, $\tilde{T}$, which is the period of firing motion in the process of pattern retrieval, is not equivalent to $T$, which is the period assumed in generating random patterns. Since no fluctuation of firing times is allowed in Eq. (17), we term the stationary state defined by (17) perfect retrieval state. In this section, we conduct the theoretical analysis to elucidate the stationary properties of the perfect retrieval state.

One purpose of the present analysis is to determine the value of the period $\tilde{T}$. In the course of the present analysis, we evaluate periodical synaptic electric currents $I_i(t) = I_{pyr,i}(t) + I_{int}(t)$ as a function of $\tilde{T}$. Once we know the time-dependent behavior of periodical synaptic electric currents $I_i(t)$, we are allowed to calculate firing motion of neurons numerically by use of Eqs. (1) and (2). Then, based on this firing motion, we determine the value of the period $\tilde{T}$ self-consistently. In what follows, $P$ is assumed to be finite since perfect retrieval is impossible with an extensive number of encoded patterns.

Substituting Eq. (16) into Eq. (4), we have

$$I_{pyr,i}(t) = \frac{A_{pyr}}{N}\sum_{\mu}\sum_{j}\tilde{W}\left\{\frac{T}{2\pi}\left(\theta_i^\mu - \theta_j^\mu\right)\right\}\sum_{k} S_{pyr}\left\{t - t_j(k)\right\}. \tag{18}$$

Then, from Eq. (17), we obtain

$$\begin{aligned} I_{pyr,i}(t) &= \frac{A_{pyr}}{N}\sum_{\mu}\sum_{j}\tilde{W}\left\{\frac{T}{2\pi}\left(\theta_i^\mu - \theta_j^\mu\right)\right\}\sum_{k=-\infty}^{\infty} S_{pyr}\left(t - \frac{\tilde{T}}{2\pi}\theta_j^1 - k\tilde{T}\right) \\ &= \frac{A_{pyr}}{N}\sum_{\mu}\sum_{j}\tilde{W}\left\{\frac{T}{2\pi}\left(\theta_i^\mu - \theta_j^\mu\right)\right\}\tilde{S}_{pyr}\left(t - \frac{\tilde{T}}{2\pi}\theta_j^1\right), \end{aligned} \tag{19}$$

where

$$\tilde{S}_{pyr}(t) = \sum_{k=-\infty}^{\infty} S_{pyr}\left(t + k\tilde{T}\right). \tag{20}$$

Substituting Eq. (5) into Eq. (20) yields the explicit form of the function $\tilde{S}_{pyr}(t)$:

$$\tilde{S}_{pyr}(t) = \frac{1}{\tau_{pyr,1} - \tau_{pyr,2}}\left(\frac{e^{-t/\tau_{pyr,1}}}{1 - e^{-\tilde{T}/\tau_{pyr,1}}} - \frac{e^{-t/\tau_{pyr,2}}}{1 - e^{-\tilde{T}/\tau_{pyr,2}}}\right), \qquad 0 \le t < \tilde{T}, \tag{21}$$

where the periodical function $\tilde{S}_{pyr}(t)$ satisfies the condition: $\tilde{S}_{pyr}(t + \tilde{T}) = \tilde{S}_{pyr}(t)$.

For the purpose of evaluating the correlation with respect to the variables $\theta_i^\mu$, we decompose Eq. (19) into the form

$$I_{pyr,i}(t) = M_i(t) + Z_i(t), \tag{22}$$

where

$$M_i(t) = \frac{A_{pyr}}{N}\sum_{j}\tilde{W}\left\{\frac{T}{2\pi}\left(\theta_i^1 - \theta_j^1\right)\right\}\tilde{S}_{pyr}\left(t - \frac{\tilde{T}}{2\pi}\theta_j^1\right), \tag{23}$$

$$Z_i(t) = A_{pyr}\sum_{\mu\ge 2}\frac{1}{N}\sum_{j}\tilde{W}\left\{\frac{T}{2\pi}\left(\theta_i^\mu - \theta_j^\mu\right)\right\}\tilde{S}_{pyr}\left(t - \frac{\tilde{T}}{2\pi}\theta_j^1\right). \tag{24}$$

Since the term $Z_i(t)$ emerges as a result of encoding a number of spatiotemporal patterns, we call the term $Z_i(t)$ crosstalk term. In the limit of $N \to \infty$, the term $M_i(t)$ is evaluated as

$$M_i(t) = \begin{cases} \dfrac{A_{pyr}}{Q}\displaystyle\sum_{q=0}^{Q-1}\tilde{W}\left\{\frac{T}{Q}\left(q_i^1 + q\right)\right\}\tilde{S}_{pyr}\left(t + \frac{\tilde{T}}{Q}q\right) & \text{finite } Q. \\ \dfrac{A_{pyr}}{2\pi}\displaystyle\int_0^{2\pi}\tilde{W}\left\{\frac{T}{2\pi}\left(\theta_i^1 + \theta\right)\right\}\tilde{S}_{pyr}\left(t + \frac{\tilde{T}}{2\pi}\theta\right)d\theta & Q \to \infty \end{cases}. \tag{25}$$

On the other hand, since $\theta_i^\mu$ $(\mu \geq 2)$ has no correlation with $\theta_i^1$, in the limit of $N \to \infty$, crosstalk term $Z_i(t)$ is evaluated as

$$Z_i(t) = A_{pyr}(P-1)\overline{W}\,\overline{S}(t) \tag{26}$$

where

$$\overline{W} = \begin{cases} \dfrac{1}{Q}\displaystyle\sum_{q=0}^{Q-1} \tilde{W}\left(\frac{T}{Q}q\right) & \text{finite } Q \\ \dfrac{1}{2\pi}\displaystyle\int_0^{2\pi} \tilde{W}\left(\frac{T}{2\pi}\theta\right) d\theta & Q \to \infty \end{cases} \tag{27}$$

and

$$\overline{S}(t) = \begin{cases} \dfrac{1}{Q}\displaystyle\sum_{q=0}^{Q-1} \tilde{S}_{pyr}\left(t + \frac{\tilde{T}}{Q}q\right) & \text{finite } Q \\ \dfrac{1}{2\pi}\displaystyle\int_0^{2\pi} \tilde{S}_{pyr}\left(\frac{\tilde{T}}{2\pi}\theta\right) d\theta & Q \to \infty \end{cases}. \tag{28}$$

Noting Eqs. (14) and (20), we obtain another representation of $\overline{W}$ and $\overline{S}(t)$:

$$\overline{W} = \begin{cases} \dfrac{1}{Q}\displaystyle\sum_{q=-\infty}^{\infty} W\left(\frac{T}{Q}q\right) & \text{finite } Q \\ \dfrac{1}{2\pi}\displaystyle\int_{-\infty}^{\infty} W\left(\frac{T}{2\pi}\theta\right) d\theta & Q \to \infty \end{cases} \tag{29}$$

and

$$\overline{S}(t) = \begin{cases} \dfrac{1}{Q}\displaystyle\sum_{q=-\infty}^{\infty} S_{pyr}\left(t + \frac{\tilde{T}}{Q}q\right) & \text{finite } Q \\ \dfrac{1}{\tilde{T}} & Q \to \infty \end{cases}, \tag{30}$$

where we use $\int_{-\infty}^{\infty} S_{pyr}(t)dt = 1$.

Following almost the same scheme as $I_{pyr,i}(t)$, from Eqs. (6) and (17), we obtain the global inhibition $I_{int}(t)$ of the form

$$I_{int}(t) = \begin{cases} \dfrac{A_{int}}{Q}\displaystyle\sum_{q=0}^{Q-1} \tilde{S}_{int}\left(t + \frac{\tilde{T}}{Q}q\right) & \text{finite } Q \\ \dfrac{A_{int}}{2\pi}\displaystyle\int_0^{2\pi} \tilde{S}_{int}\left(t + \frac{\tilde{T}}{2\pi}\theta\right) d\theta & Q \to \infty \end{cases}, \tag{31}$$

where

$$\tilde{S}_{int}(t) = \sum_{k=-\infty}^{\infty} S_{int}\left(t + k\tilde{T}\right). \tag{32}$$

Substituting Eq. (6) into Eq. (32), we obtain the explicit form of the function $\tilde{S}_{int}(t)$:

$$\tilde{S}_{int}(t) = \frac{-1}{\tau_{int,1} - \tau_{int,2}} \left( \frac{e^{-t/\tau_{int,1}}}{1 - e^{-\tilde{T}/\tau_{int,1}}} - \frac{e^{-t/\tau_{int,2}}}{1 - e^{-\tilde{T}/\tau_{int,2}}} \right), \qquad 0 \leq t < \tilde{T}, \tag{33}$$

where the periodical function $\tilde{S}_{int}(t)$ satisfies the condition: $\tilde{S}_{int}(t+\tilde{T}) = \tilde{S}_{int}(t)$. Utilizing $\int_{-\infty}^{\infty} S_{int}(t)dt = -1$, we obtain another representation of $I_{int}(t)$:

$$I_{int}(t) = \begin{cases} \dfrac{A_{int}}{Q}\displaystyle\sum_{q=-\infty}^{\infty} S_{int}\left(t + \frac{\tilde{T}}{Q}q\right) & \text{finite } Q \\ -\dfrac{A_{int}}{\tilde{T}} & Q \to \infty \end{cases}. \tag{34}$$

Substituting Eqs. (22) and (26) into Eq. (3), we obtain the periodical synaptic electric currents $I_i(t)$ of the form

$$I_i(t) = M_i(t) + I_{int}(t) + A_{pyr}(P-1)\overline{W}\,\overline{S}(t). \tag{35}$$

Note that now all the terms in the right hand side of Eq. (35) are evaluated as a function of $\theta_i^1$ and $\tilde{T}$, and hence we can evaluate the time-dependent behavior of $N$ neurons as a function $\theta_i^1$ and $\tilde{T}$ based on the dynamics (1) and (2) together with Eq. (35). In the case of perfect retrieval, the periodical behavior of every neuron is identical, but shifts with respect to time depending on the value $\theta_i^1$. In fact, noting Eq. (25), one can show that a sum of synaptic electric currents $I_i(t)$ satisfies the condition:

$$I_i\left(t+\frac{\tilde{T}}{2\pi}\theta_i^1\right) \quad = \quad I_j\left(t+\frac{\tilde{T}}{2\pi}\theta_j^1\right), \qquad i=1,\ldots,N,\ j=1,\ldots,N. \tag{36}$$

By use of this property, the behavior of $N$ neurons is easily evaluated once we know the behavior of a single neuron, that is, what we need to solve is not a many-body problem but a single-body problem. We hence focus on investigating the behavior of a single neuron with $\theta_i^1=0$ in what follows.

For a single neuron with $\theta_i^1=0$, we rewrite the dynamics Eqs. (1), (2) and (35) in the form

$$\dot{V} \quad = \quad f(V, W_1, \ldots, W_n) + I(t), \tag{37}$$

$$\dot{W}_j \quad = \quad g_j(V, W_1, \ldots, W_n), \qquad j=1,\ldots,n \tag{38}$$

with

$$I(t) = M(t) + I_{int}(t) + A_{pyr}(P-1)\overline{W}\,\overline{S}(t), \tag{39}$$

where, from Eq. (25), the term $M(t)$ is rewritten in the form

$$M(t) = \begin{cases} \dfrac{A_{pyr}}{Q}\displaystyle\sum_{q=0}^{Q-1}\tilde{W}\left(\frac{T}{Q}q\right)\tilde{S}_{pyr}\left(t+\frac{\tilde{T}}{Q}q\right) & \text{finite } Q. \\ \dfrac{A_{pyr}}{2\pi}\displaystyle\int_0^{2\pi}\tilde{W}\left(\frac{T}{2\pi}\theta\right)\tilde{S}_{pyr}\left(t+\frac{\tilde{T}}{2\pi}\theta\right)d\theta & Q\to\infty \end{cases}. \tag{40}$$

Note that $I_{int}(t)$, $\overline{W}$, and $\overline{S}(t)$ in Eq. (39) is given by Eqs. (31),(27), and (28) respectively.

By use of the Hodgkin-Huxley equations, we can evaluate the dynamics (37)-(39) numerically for arbitrary value of $\tilde{T}$. As is describe in Fig. 2, with $\tilde{T}$ that is sufficiently close to the solution $\tilde{T}^*$, the neuron exhibits periodical firing behavior, and hence the firing times are written in the form

$$t(k) = r\left(\tilde{T}\right) + k\tilde{T}, \qquad k=\ldots,-2,-1,0,1,2.\ldots. \tag{41}$$

Note that we can evaluate the explicit form of the function $r(\tilde{T})$ by conducting the numerical integration of the dynamics (37)-(39) for various value of $\tilde{T}$.

On the other hand, since we evaluate the periodical synaptic electric current (39) based on the assumption (17), with the solution $\tilde{T}^*$ the firing times take the form

$$t(k) = \frac{\tilde{T}^*}{2\pi}\theta + k\tilde{T}^* = k\tilde{T}^* \qquad k=\ldots,-2,-1,0,1,2.\ldots. \tag{42}$$

Hence, from Eqs. (41) and (42), we obtain the condition

$$r(\tilde{T}^*) = 0. \tag{43}$$

Since we have evaluated the explicit form of the function $r(\tilde{T})$ by the numerical integration, we can solve Eq. (43) so as to obtain the solution $\tilde{T}^*$.

In the above analysis, we did not take account of the stability of the solution. Strictly speaking, the present networks may happen to fail in perfect retrieval for lack of the stability even when $\tilde{T}$ is successfully evaluated in the above analysis. However, as far as we investigate by numerical simulations, every solution we obtain in the present analysis seems to ensure the stability as will be shown in section 4 and 5.

### 3.1 The time window $W(\Delta t)$ with the negative and positive parts is suitable to encode a number of spatiotemporal patterns

It has been shown that the crosstalk term of the standard type of the Hopfield model vanishes with an appropriate learning rule as far as the number of encoded patterns is finite. In such a case, perfect retrieval is always realized irrespective of the number of encoded patterns. In the case of the present neural network, the periodical synaptic electric current (35) includes the crosstalk term $A_{pyr}(P-1)\overline{W}\,\overline{S}(t)$, which is proportional to $P-1$. From Eqs. (5) and (30), one can show that $\overline{S}(t)$ always takes the positive value. Therefore, when $\overline{W}$ takes the nonzero value, the quality of pattern retrieval changes depending on the number of encoded patterns; As $P$ increases, the size of the synaptic electric current $I_i(t)$ increases or decreases depending on the sign of $\overline{W}$, and eventually perfect retrieval becomes impossible. For these reasons, it may be highly desirable that $\overline{W}$ takes the small value in the present purpose.

It should be noted that *the quantity $\overline{W}$, which is defined by Eq. (29), is the average of the function $W(\Delta t)$ over the time $\Delta t \in (-\infty, \infty)$, and thus the presence of the negative and positive parts in the form of the time window $W(\Delta t)$ is of advantage to reduce the value of $\overline{W}$.* In fact, the form of the time window $W(\Delta t)$ we assume in the present study satisfies the condition:

$$\begin{cases} \displaystyle\sum_{q=-\infty}^{\infty} W\left(\frac{T}{Q}q\right) = 0 & \text{finite } Q \\ \displaystyle\int_{-\infty}^{\infty} W(\tau)\, d\tau = 0 & Q \to \infty \end{cases} . \tag{44}$$

Therefore, from Eq. (29), we obtain

$$\overline{W} = 0. \tag{45}$$

In the present case, the crosstalk term $A_{pyr}(P-1)\overline{W}\,\overline{S}(t)$ vanishes completely. Accordingly, perfect retrieval is always realized irrespective of the number of encoded patterns as far as $P$ is finite. In what follows, setting $\overline{W} = 0$, we analytically evaluate the stationary behavior of the network. It turns out that the result of the present analysis shows the good agreement with the results of the numerical simulations even when a number of patterns are encoded.

# 4 The case of continuous type of patterns ($Q \to \infty$)

## 4.1 Perfect retrieval with the weak intensity of global inhibition

For the initial condition of the network, we set state of neurons $(V_i, \{W_i\})$ $(i = 1, \ldots, N)$ to be at the stable fixed point of the dynamics of Eqs. (1) and (2) with $I_i(t) = 0$. Since all neurons keep quiescent without any external stimuli, to invoke initial firings that act as a trigger to retrieve the target pattern $s_i^1$, we use the external electric current $I_{ext,i}(t)$ of the form

$$I_{ext,i}(t) = \begin{cases} A_{ext} & 0 \le \tilde{s}_i^1 < a_{ext}T_{ext} \text{ and } \tilde{s}_i^1 \le t < \tilde{s}_i^1 + \Delta t_{ext} \\ 0 & \text{otherwise} \end{cases} \tag{46}$$

with

$$\tilde{s}_i^1 = \frac{T_{ext}}{2\pi}\theta_i^1, \tag{47}$$

where the parameters $A_{ext}, T_{ext}, \Delta t_{ext},$ and $a_{ext}$ are appropriately chosen so that an initial part of the target pattern is forced to be retrieved. In what follows, we set $A_{ext} = 10, \Delta t_{ext} = 1, a_{ext} = 0.2,$ and $T_{ext} \sim \tilde{T}$. (Once note that $\tilde{T}$ is not equivalent to $T$.)

In Fig. 3(**a**), we describe the result of the numerical simulation with the weak intensity of global inhibition $A_{int} = 250$ in the case of $Q \to \infty$ and $A_{pyr} = 20000$. After the initial firings that are invoked by the application of the external electric current $I_{ext,i}(t)$, perfect retrieval is realized as a result of the emergence of the periodical synaptic electric currents. Since it is somewhat difficult to see whether the target pattern is retrieved or not in Fig. 3(**a**), setting the vertical axis to represent the phase variables of the target pattern $\theta_i^1$, we replot the same result in Fig. 3(**b**), where we see the continuous type of firing pattern implying the occurrence of the perfect retrieval of the target pattern.

In Fig. 4(**a**), we describe the dynamical behavior of a neuron with $\theta_i^1 = 0$. In this result, the neuron is found to fire periodically after a long time. In the case of perfect retrieval, we are allowed to apply the theoretical analysis conducted above so as to evaluate the periodical firing motion of a neuron in the limit of an infinite number of neurons. The result of the theoretical analysis is described in Fig. 4(**b**). Good agreement between the numerical result in Fig. 4(**a**) and the theoretical result in Fig. 4(**b**) implies the validity of the present analysis.

## 4.2 The phase transition occurs with change of the intensity of global inhibition

As is discussed in the previous subsection, in the case of the weak intensity of global inhibition, perfect retrieval of continuous type of patterns is realized. On the other hands, when the strong intensity of global inhibition is applied, we observe the discrete type of firing pattern as is described in Fig. 5, where we set $A_{int} = 1250$. In the present study, we call such the discrete type firing pattern the gamma oscillation. In this discrete type of firing pattern we find a number of components of the continuous type of firing patterns of short duration. During the occurrence of each component, inhibitory synaptic electric currents $I_{int}(t)$ accumulate until they become to suppress more firings of neurons. After a stop of continuous firings, the inhibitory synaptic electric currents $I_{int}(t)$ decay fast owing to the short decay times $\tau_{int,1}$ and $\tau_{int,2}$. Subsequently, neurons begin to fire again because of the synaptic electric currents $I_{pyr,i}(t)$, which have the longer decay times $\tau_{pyr,1}$ and $\tau_{pyr,2}$ than those of $I_{int}(t)$. The gamma oscillation in the present study is induced by the iteration of this process.

As far as perfect retrieval is concerned, it is uncomplicated to investigate the properties of the stationary state analytically, while the analysis becomes quite difficult once the system settles into the other state such as discrete type of firing patterns. Nevertheless, within the scope of the present analysis, we can determine the critical intensity of global inhibition $A^c_{int}$, which characterizes the phase transition between the perfect retrieval state and the other state. In Fig. 6, we depict the $A_{int} - A_{pyr}$ phase diagram showing the condition for the occurrence of perfect retrieval. In the region denoted by PR, the period of the perfect retrieval $\tilde{T}$ is successfully evaluated, that is, the perfect retrieval is realized in the region PR, while outside the region the perfect retrieval is impossible.

In this phase diagram, the critical intensity of the global inhibition in the case of $A_{pyr} = 20000$ is evaluated as $A^c_{int} \sim 630$. To clarify the occurrence of the phase transition at this critical intensity $A^c_{int}$, for various value of $A_{int}$ we compute a distribution of the inter spike intervals (ISIs), which are the time intervals of sequential firings of neurons in the numerical simulations. Note that what we compute is not the ISIs of a single neuron but the ISIs of all neurons, that is, when neuron $i$ and neuron $j$ fire sequentially at $t_i$ and $t_j$ respectively, we compute the time interval $t_j - t_i$. The result of the computation of ISIs is plotted in Fig. 7(**a**). Since, below the critical intensity $A^c_{int}$, the continuous type of firing pattern is realized, every ISI becomes almost zero. Beyond the critical intensity $A^c_{int}$, owing to the occurrence of non-perfect retrieval we see the distribution of the ISIs with two components, namely, the component with the short ISIs and the component with the long ISIs. The appearance of the component with the short ISIs is attributed to the emergence of the continuous type of firing patterns of short duration while the appearance of the long ISIs is attributed to the period during which firing of neurons are suppressed. In Fig. 7(**a**), we clearly see the occurrence of the phase transition at the critical intensity of global inhibition $A^c_{int}$.

# 5 The case of discrete type of patterns ($Q = 10$)

As in the case of continuous type of patterns, perfect retrieval is realized even in the case of discrete type of patterns. In Fig. 9, we describe the result of the numerical simulations, where we see the discrete type of firing pattern as a result of the retrieval of the discrete type of pattern with $Q = 10$. In the numerical simulation in Fig. 9, we assume the weak intensity of global inhibition ($A_{int} = 250$). In the case of discrete type of patterns, with change of the intensity of global inhibition we find a variety of stationary behavior, which is much richer than that in the case of continuous type of patterns.

## 5.1 Two types of perfect retrieval state

Following the same scheme as continuous type of patterns, we describe the $A_{int} - A_{pyr}$ phase diagram in Fig. 8. Unlike the case of the continuous type of patterns, we find the two kinds of critical intensity of global inhibition $A^c_{int}(1)$ and $A^c_{int}(2)$ in the region with $13000 \lesssim A_{pyr} \lesssim 23000$, that is, we see two types of the phase transitions with change of the intensity of global inhibition $A_{int}$.

To elucidate the nature of these two types of the phase transitions, fixing $A_{pyr} = 17000$, we conduct the numerical simulations for the various value of $A_{int}$. In Fig. 9, we describe the result of the numerical simulations with $A_{int} = 250$, which is weaker than the first critical intensity $A^c_{int}(1)$. As is expected from the phase diagram in Fig. 8, perfect retrieval occurs with this intensity of global inhibition. In this case, global inhibition is so weak that the influence of global inhibition on the properties of the retrieval state is insignificant.

On the other hand, when we apply the stronger intensity of global inhibition $A_{int}$ than the second critical intensity $A^c_{int}(2)$, the global inhibition exerts the significant influence on the nature of the retrieval process. In Fig. 10, we describe the result of numerical simulation with $A_{int} = 1250$, where we see the perfect retrieval with the long period. With the strong intensity of global inhibition the gamma oscillation arises and affects to make the retrieval period long. In the retrieval process of discrete type of patterns, a cluster of neurons with $\theta_i^1 = 2\pi\frac{1}{Q}$ fire after a cluster of neurons with $\theta_i^1 = 2\pi\frac{0}{Q}$ fire. Firing of a cluster of neurons with $\theta_i^1 = 2\pi\frac{0}{Q}$ induces two types of synaptic electric currents: the global inhibition as well as the excitatory synaptic electric current that evokes firing of a cluster of neurons with $\theta_i^1 = 2\pi\frac{1}{Q}$. The emergence of the global inhibition prevents the immediate firing of the next cluster. After a certain time interval the global inhibition decays, and then a cluster of neurons with $\theta_i^1 = 2\pi\frac{1}{Q}$ begins to fire owing to the excitatory synaptic electric current. As a result of the interaction of these processes, the gamma oscillation arises so that the pattern retrieval occurs with the long period. It turns out that the occurrence of the gamma oscillation gives the appropriate spike timings for memory retrieval of discrete type of patterns though it is of disadvantage in the case of continuous type of patterns.

### 5.2 Disordered state with the intermediate intensity of global inhibition

With the intermediate intensity of global inhibition $A_{int}^c(1) < A_{int} < A_{int}^c(2)$ the perfect retrieval is impossible as is shown in Fig. 8. In Fig. 11, we describe the result of the numerical simulations with the intermediate intensity of global inhibition $A_{int} = 750$, where we find the disordered firing pattern. In this case, the dynamical behavior of a neuron is so complicated that it is quite difficult to specify whether the time evolution of $I_i(t)$ and $V_i$ is periodic or not. In addition, neurons with the same value of $\theta_i^1$ exhibits the different dynamical behavior, because we see the slight distribution of the firing times of neurons with the same value of $\theta_i^1$ in Fig. 11(**a**).

For the purpose of elucidating the difference of the disordered state from the prefect retrieval state, we compute the ISIs for various value of $A_{int}$ as is described in Fig. 12. In the two intervals $A_{int} \leq A_{int}^c(1)$ and $A_{int}^c(2) \leq A_{int}$, where the perfect retrieval is expected to occur, the ISIs take 0 or $\tilde{T}/Q$. Analytically evaluated $A_{int}$-dependence of $\tilde{T}/Q$ in Fig. 12(**b**) shows the good agreement with the result of numerical simulations in Fig. 12(**a**). Meanwhile, in the interval $A_{int}^c(1) < A_{int} < A_{int}^c(2)$, where perfect retrieval is impossible, we see the quite complicated distribution of the ISIs. It turns out that with change of the intensity of global inhibition two types of phase transitions occur at the critical intensity $A_{int}^c(1)$ and $A_{int}^c(2)$.

## 6 Discussion

We have investigated associative memory neural networks of spiking neurons interacting through two types of synaptic electric currents: currents via plastic synapses and global inhibition. Based on the result of the electrophysiological experiments, we have introduced the spike-timing-dependent learning rule (16), which encodes spike timings of neurons so that networks function as associative memory.

To elucidate the stationary properties of perfect retrieval state, we have evaluated the periodical firing motion of neurons analytically in the limit of an infinite number of neurons. Based on this method of analysis, we have shown that the present form of the time window $W(\Delta t)$ has the great advantage in encoding a number of spatiotemporal patterns since the crosstalk term is proportional to the quantity $\overline{W}$, which has been shown to vanish owing to the negative and positive parts of the time window $W(\Delta t)$.

We have examined to encode two types of spatiotemporal patterns: continuous type of patterns ($Q \to \infty$) and discrete type of pattern ($Q = 10$). In the case of continuous type of patterns, perfect retrieval is realized with the weak intensity of global inhibition, while it is impossible with the strong intensity of global inhibition since the occurrence of gamma oscillation prevents the realization of perfect retrieval. Applying the present method of analysis we have drawn the $A_{int} - A_{pyr}$ phase diagram in Fig. 6, in which we have evaluated the critical intensity $A_{int}^c$ characterizing the phase transition between the perfect retrieval state and the the other state.

Meanwhile, in the case of discrete type of patterns, we have found two types of perfect retrieval state, which is characterized by the two critical intensity: $A_{int}^c(1)$ and $A_{int}^c(2)$ (see the $A_{int} - A_{pyr}$ phase diagram in Fig. 8). When the intensity of global inhibition $A_{int}$ is weaker than the first critical intensity $A_{int}(1)$, retrieval is successful as in the case of the continuous type of patterns. In addition, even with the strong intensity of global inhibition $A_{int}$, perfect retrieval is realized provided that the intensity $A_{int}$ is stronger than the second critical intensity $A^c_{int}(2)$. In this case, the gamma oscillation arises, and it gives the appropriate spike timings for retrieval of discrete type of patterns. With the intermediate intensity $A^c_{int}(1) < A_{int} < A^c_{int}(2)$ we have observed the rather complicate firing patterns as is shown in Fig. 11.

It is noted that the crosstalk term $A_{pyr}(P-1)\overline{W}\,\overline{S}(t)$ vanishes if the time window $W(\Delta t)$ satisfies the condition (44). Although the results of some electrophysiological experiments indicate the slight dominance of the positive part of the time window $W(\Delta t)$ over the negative part, the presence of the negative part is still profitable to reduces the size of $\overline{W}$ because of Eq. (29). Although the function of the form like the Mexican hat may also satisfy the condition (44), the present form of the time window $W(\Delta t)$ is considered to be more adequate to encode spatiotemporal patterns since the emergence of the excitatory synaptic electric current before firing of a neuron in the retrieval process is attributed to the positive part of the the time window $W(\Delta t)$, while the fast decay of this excitatory synaptic electric current after firing is attributed to the negative part. We have to, however, keep in mind that the assumption that synaptic efficacy $J_{ij}$ can take negative value as well as positive value may be somewhat implausible from a biological point of view since synapses among pyramidal cells are found to be excitatory in experiments. The present learning rule (16), which gives either negative synaptic efficacy or positive synaptic efficacy by chance, is introduced based on the rough estimation of the modification of synaptic efficacy in the subsection 2.1. This rough estimation is somewhat tricky since some quantities diverge in its procedure owing to the absence of the dumping effect. Rubin et al. has investigated the modification of synaptic efficacy incorporating several types of dumping scheme so that the synaptic efficacy is restricted to positive value [60], and such the approach may be required to get further insight.

In the present study, we have investigated the stationary properties of perfect retrieval state analytically in the limit of an infinite number of spiking neurons provided that the number of encoded patterns is finite. The method of our analysis is extended to the cases such as superimposed firing patterns, in which the firing times of neurons are defined as

$$t_i(k) = \frac{\tilde{T}}{2\pi/l}\left\{\theta_i^1 \bmod (2\pi/l)\right\} + k\tilde{T}, \qquad i = 1,\ldots,N,\ k = \ldots,-2,-1,0,1,2,\ldots, \tag{48}$$

where pattern 1 is the target pattern and positive integer $l$ denotes the degree of superimposing (Taking $l = 1$ corresponds to the case of perfect retrieval). In addition, more complicated firing patterns such as a mixture state, in which two or more patterns are retrieved at the same time, are expected to be realized under an appropriate initial condition, though theoretical treatment of them may be difficult to achieve. We have shown that perfect retrieval is realized in the region represented by PR in Figs. 6(**a**) and 8(**a**). Nevertheless, what happens below the perfect retrieval phase remains unclear for lack of the method of analysis. On the basis of the numerical simulations, we have shown that disordered state can occur below the perfect retrieval phase in the case of discrete type of patterns as is described in Fig. 11. Whether this disordered firing patterns is chaotic or not is of interest, but is beyond the scope of the present study.

It is worth noting that the learning rule (10) is applicable to a wide class of spatiotemporal patterns. For example, following almost the same scheme as the present study, we would be able to encode spike trains generated by independent Poisson process. In this case, the firing rate assumed in the Poisson process is expected to affect the quality of memory retrieval, because the presence of the refractory period of neurons prevents retrieval of the spike trains with the high firing rate. It is of interest to investigate the properties of the retrieval process of the present model under the influence of white noise. It has turned out that the occurrence of the gamma oscillation contributes to the realization of retrieval of discrete type of patterns, and investigating the stability of such the gamma oscillation against noise is particular interesting. It is also of interest to study the retrieval process of networks of neurons with heterogeneity.

It seems to be difficult to carry out the rigorous derivation of the storage capacity of the present model though it might be possible to evaluate approximate value of the storage capacity by reducing the present model into networks of simple phase oscillators [61]. In the previous study we have introduced the method to reduce networks of spiking neurons into the Hopfield models when networks of spiking neurons exhibits roughly synchronized firing [23]. This reduction technique might be also applicable to the present model to obtain the approximate value of the storage capacity when discrete type of patterns are encoded and firing pattern of neurons becomes discrete. For the purpose of elucidating how global inhibition affect the retrieval properties of the network, we numerically estimate the storage capacity $\alpha^c = P^c/N$ for various value of global inhibition in the case of the discrete type of patterns ($Q = 10, A_{pyr} = 17000$, and $N = 2000$, see Fig. 13). With the strong intensity of global inhibition ($A_{int} = 1250$), the storage capacity is estimated to be $\alpha^c \sim 0.008$, while with the weak intensity of global inhibition ($A_{int} = 250$) it is estimated to be $\alpha^c \sim 0.006$. It seems that networks with strong intensity of global inhibition is a little more tolerant with regard to such fluctuation as crosstalk term than that with weak intensity of global inhibition.

In the present study, the occurrence of the gamma oscillation is assumed for the purpose of controlling spike timings of neurons. On the other hand, it is also possible to control spike timings of neurons by assuming the conduction delay with respect to action potentials. In such a case, the spike-timing-dependent learning rule takes the form

$$J_{ij} = \frac{1}{N} \sum_{\mu=1}^{P} \tilde{W}\left(s_i^\mu - s_j^\mu - d_{ij}\right), \tag{49}$$

where $d_{ij}$ represents the conduction delay of action potential from neuron $j$ to neuron $i$. Even in this case, following the same scheme as [23], the stationary properties of perfect retrieval state can be evaluated analytically provided that the time delays $d_{ij}$ are independent random variables obeying a certain probability distribution $P_d(d_{ij})$.

Finally, we discuss the implication of the present study in the light of the experimental studies regarding place cells in the rat hippocampus. It has been reported that place cells in the rat hippocampus becomes to exhibit the environment-specific distribution of center of place field after exploring several environments (i.e., exploring a number of test circuits) [62–64]. These results imply that the rat hippocampus is capable of memorizing not only a single pattern but also a number of patterns, and this aspect of the rat hippocampus may be well accounted for by the present model. In the present study, encoded periodical spatiotemporal patterns are retrieved with the different time scale depending on the intensity of global inhibition. Some recent results of experiments begin to suggest that the spike sequences observed in the hippocampus of running rats is replayed in a time-compressed manner during sharp wave burst in slow-wave sleep [21, 65]. These results imply that the spike sequences memorized in a running rat is replayed with the different time scale when rats is in slow-wave sleep. It may be possible to give some qualitative explanation to this phenomenon if we observe the retrieval period of more precise neural network models changing the value of some parameters such as the intensity of global inhibition. For a more complete understanding of a real nervous system it might be necessary to assume interactions among interneurons though we neglect them for brevity in the present study.

When a rat is running, a population of neurons in the hippocampus exhibit the theta rhythm, which is roughly synchronized firings of neurons with a characteristic frequency 7-9Hz. To get more insight into the information processing conducted in the hippocampus it may be necessary to pay more attention to the role of the theta rhythm in the retrieval process of the spatiotemporal patterns. In the presence of the theta rhythm, the sequential firings of neurons are suppressed during the period when the averaged activity of neurons takes the low value. In such a case, we may need to assume some kind of synaptic electric currents that act as a trigger to retrieve the target pattern for individual theta cycles. Giving a good account for these problems will be one of the future targets of our study.

# Acknowledgements

We gratefully acknowledge helpful discussions with Prof. Fukai and Prof. Yamaguchi on several points in the present paper.

# Appendix

# A The Hodgkin-Huxley equations

Hodgkin-Huxley equations are the ordinary differential equations with four degrees of freedom, which have been developed to describe the spike generation of the squid's giant axon [57]. In the present study, for the dynamics $f(V, W_1, \ldots, W_n)$ and $g_j(V, W_1, \ldots, W_n)$ ($j =$

$1, \ldots, n$), we assume the Hodgkin-Huxley equations, which are written in the form

$$
\begin{aligned}
C_m \, f\,(V, W_1, \ldots, W_3) &= \overline{g}_{Na} W_2^3 W_1 \,(V_{Na} - V) + \overline{g}_K W_3^4 \,(V_K - V) \\
&\quad + \overline{g}_L \,(V_L - V)\,, && (50) \\
g_1\,(V, W_1, \ldots, W_3) &= \alpha_1\,(1 - W_1) - \beta_1 W_1, && (51) \\
g_2\,(V, W_1, \ldots, W_3) &= \alpha_2\,(1 - W_2) - \beta_2 W_2, && (52) \\
g_3\,(V, W_1, \ldots, W_3) &= \alpha_3\,(1 - W_3) - \beta_3 W_3 && (53)
\end{aligned}
$$

with

$$
\begin{aligned}
\alpha_1 &= 0.01\,(10 - V) \bigg/ \left\{ \exp\left(\frac{10 - V}{10}\right) - 1 \right\}, && (54) \\
\beta_1 &= 0.125 \exp\left(-V/80\right), && (55) \\
\alpha_2 &= 0.1\,(25 - V) \bigg/ \left\{ \exp\left(\frac{25 - V}{10}\right) - 1 \right\}, && (56) \\
\beta_2 &= 4 \exp\left(-V/18\right), && (57) \\
\alpha_3 &= 0.07 \exp\left(-V/20\right), && (58) \\
\beta_3 &= 1 \bigg/ \left\{ \exp\left(\frac{30 - V}{10}\right) - 1 \right\}, && (59)
\end{aligned}
$$

where $V$ represents the membrane potential, and $W_1$ and $W_2$ the activation and inactivation variables of the sodium current, and $W_3$ the activation variable of the potassium current. The values of parameters are $V_{Na} = 50$ (mV), $V_K = -77$ (mV), $V_L = -54.4$ (mV), $\overline{g}_{Na} = 120$ (mS/cm$^2$), $\overline{g}_K = 36$ (mS/cm$^2$), $\overline{g}_L = 0.3$ (mS/cm$^2$), and $C_m = 1$ ($\mu$F/cm$^2$).

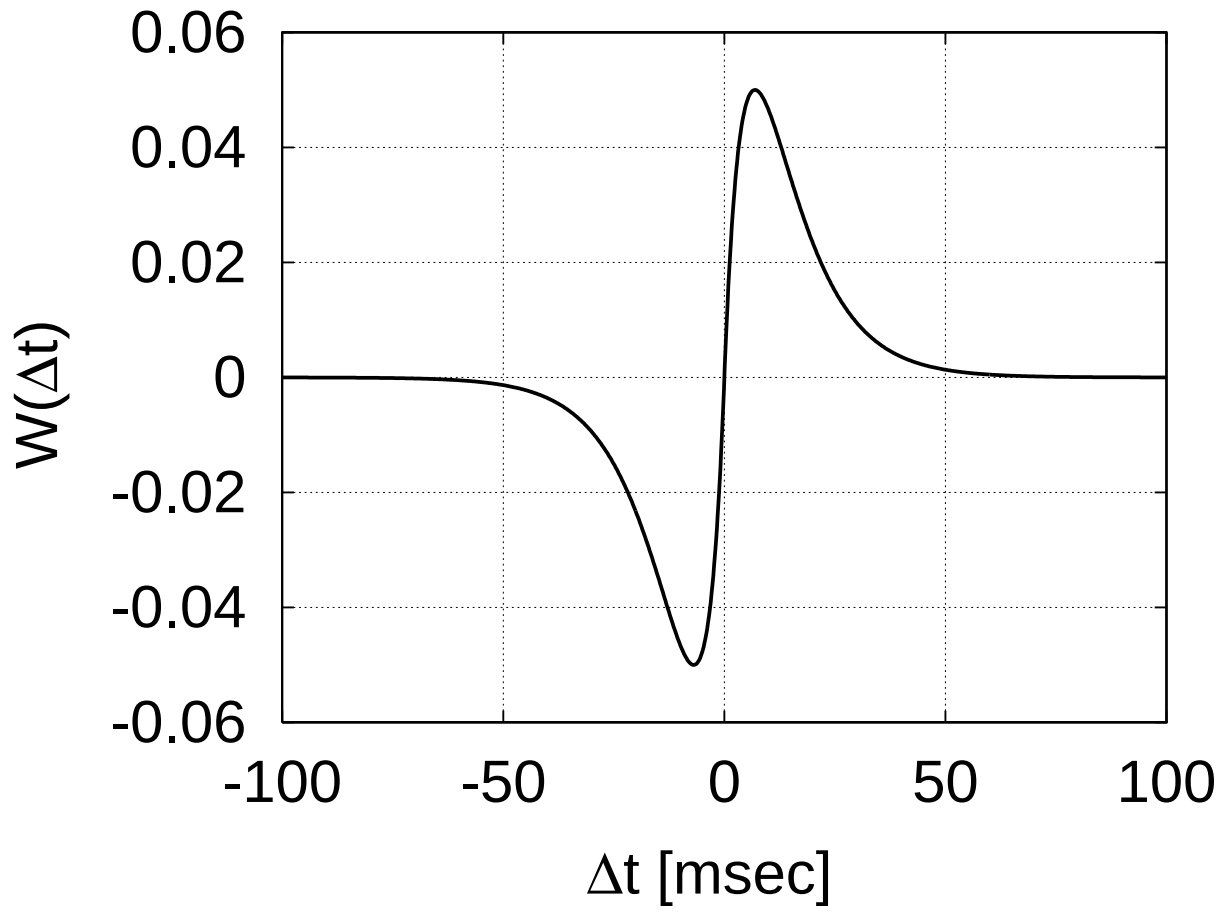

Figure 1: The shape of the time window $W(\Delta t)$ with $\tau_{W,1} = 10$ (msec) and $\tau_{W,2} = 5$ (msec).

(**a**)

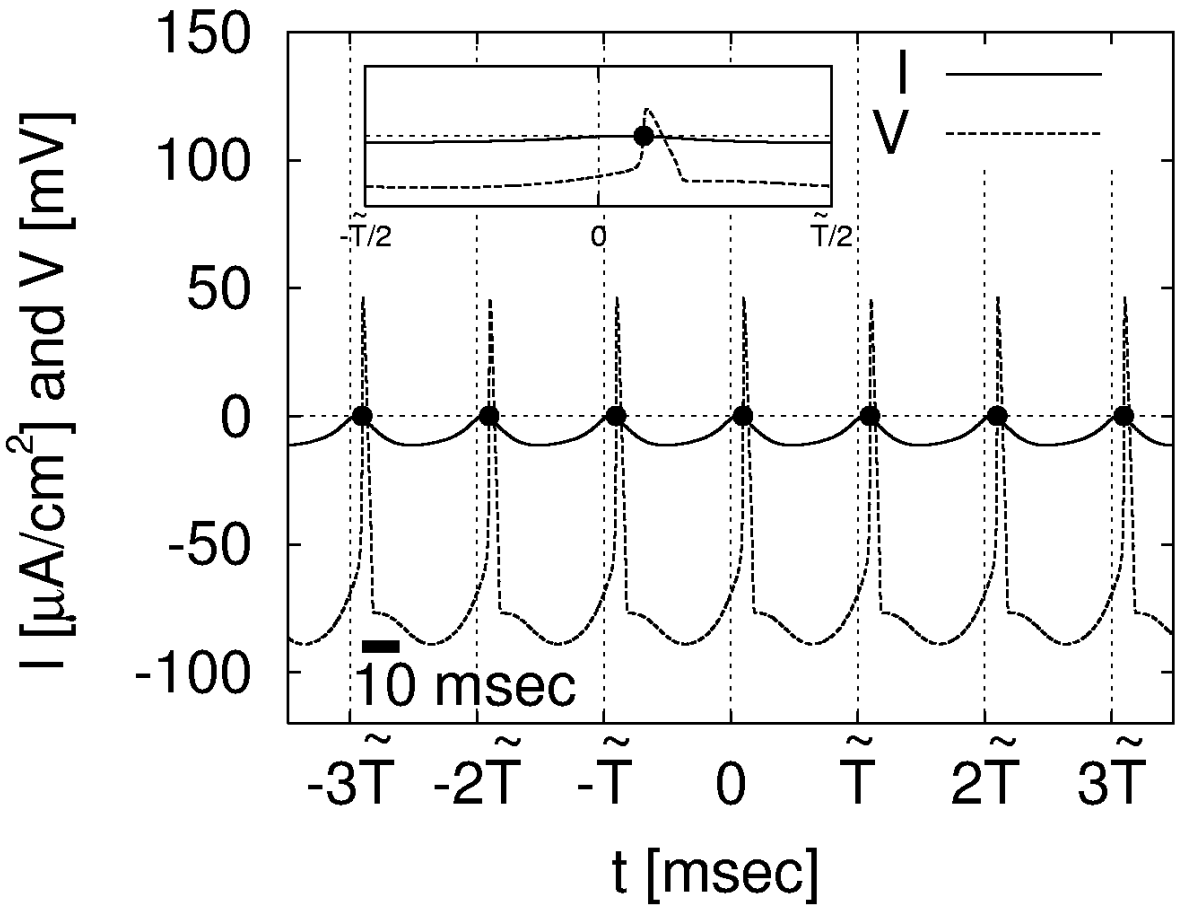

(**b**)

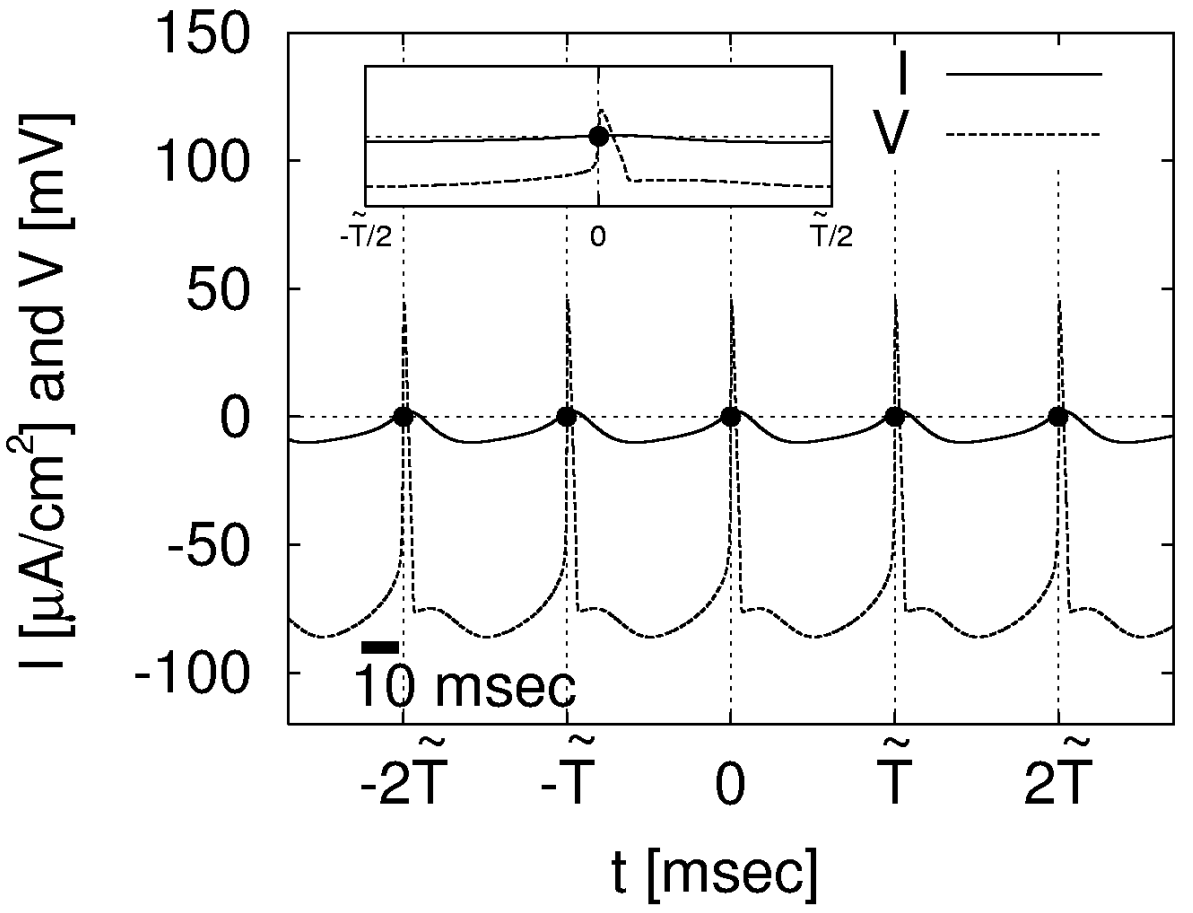

(**c**)

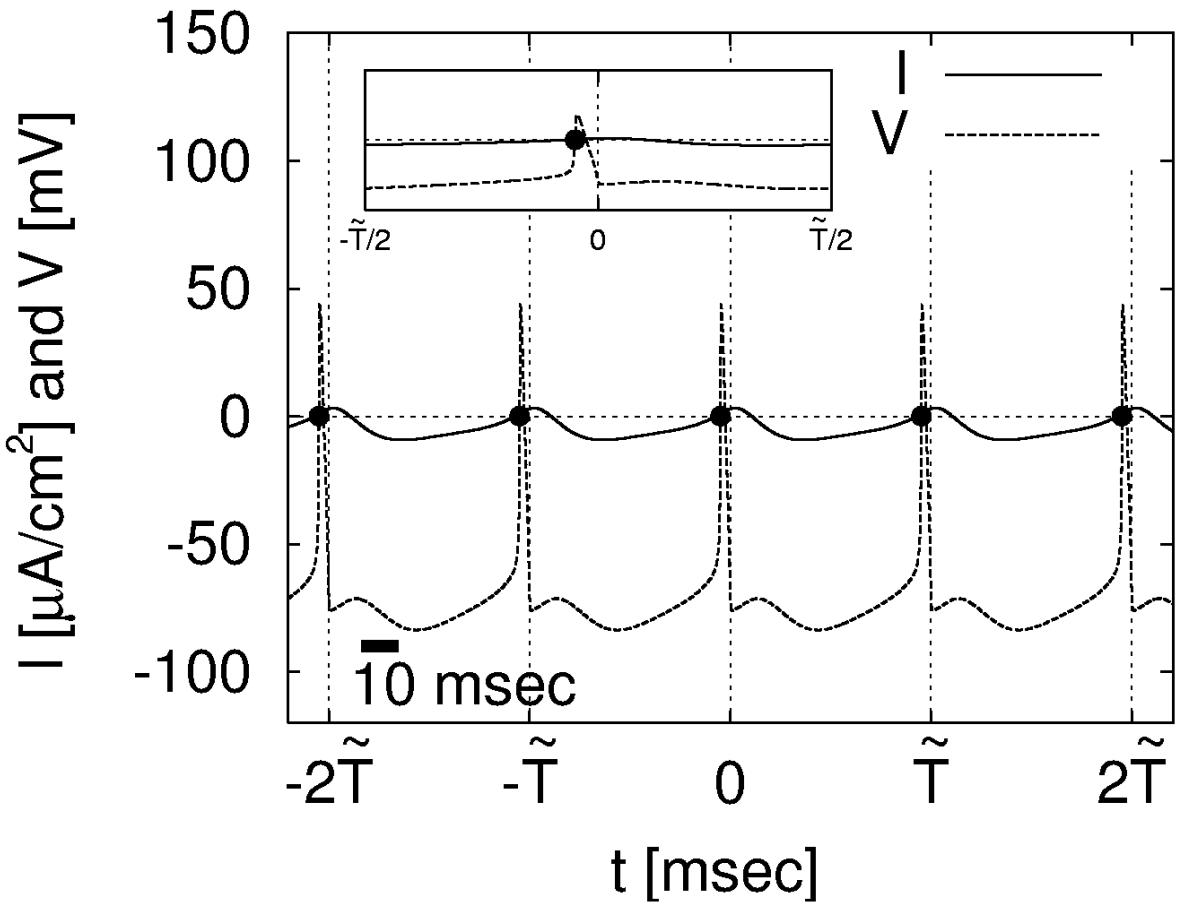

Figure 2: The stationary behavior of the single neuron dynamics (37)-(39), which is analytically derived for the purpose of evaluating the periodical firing behavior of a network of neurons. For the several value of $\tilde{T}$, which is close to the solution $\tilde{T}^*$, the time evolution of $I(t)$ and $V$ are plotted together with the firing times, which are marked by closed circles ((**a**) $\tilde{T} = \tilde{T}^* - 10$, (**b**) $\tilde{T} = \tilde{T}^*$, (**c**) $\tilde{T} = \tilde{T}^* + 10$). Inset, A magnification representing the behavior of $I(t)$ and $V$ within one period. Note that the firing time takes the form $t(k) = k\tilde{T}$ $(k = \ldots, -2, -1, 0, 1, 2, \ldots)$ only in the case of (**b**). The value of parameters are $Q \to \infty, A_{pyr} = 20000$, and $A_{int} = 250$, which are the same values as we use in Figs. 3 and 4.

(**a**)

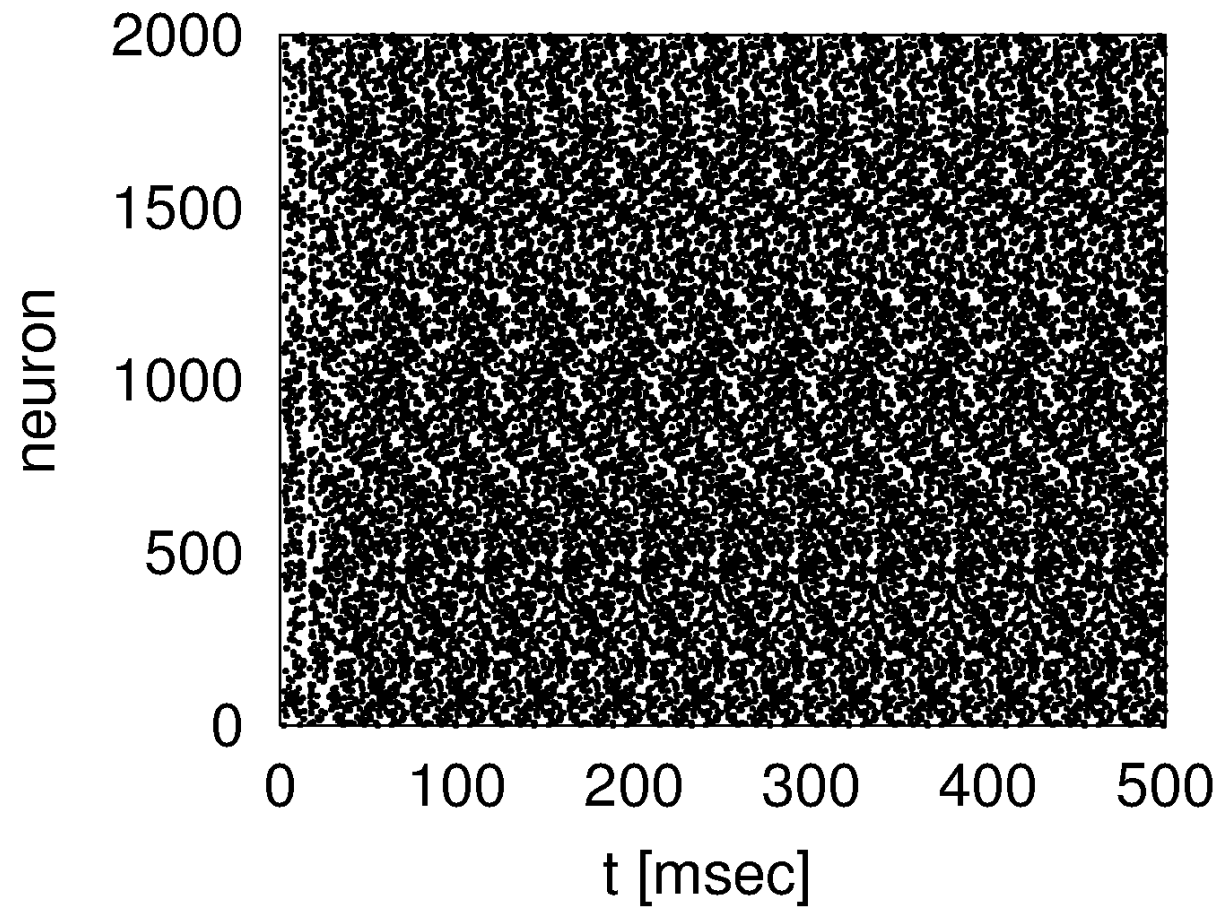

(**b**)

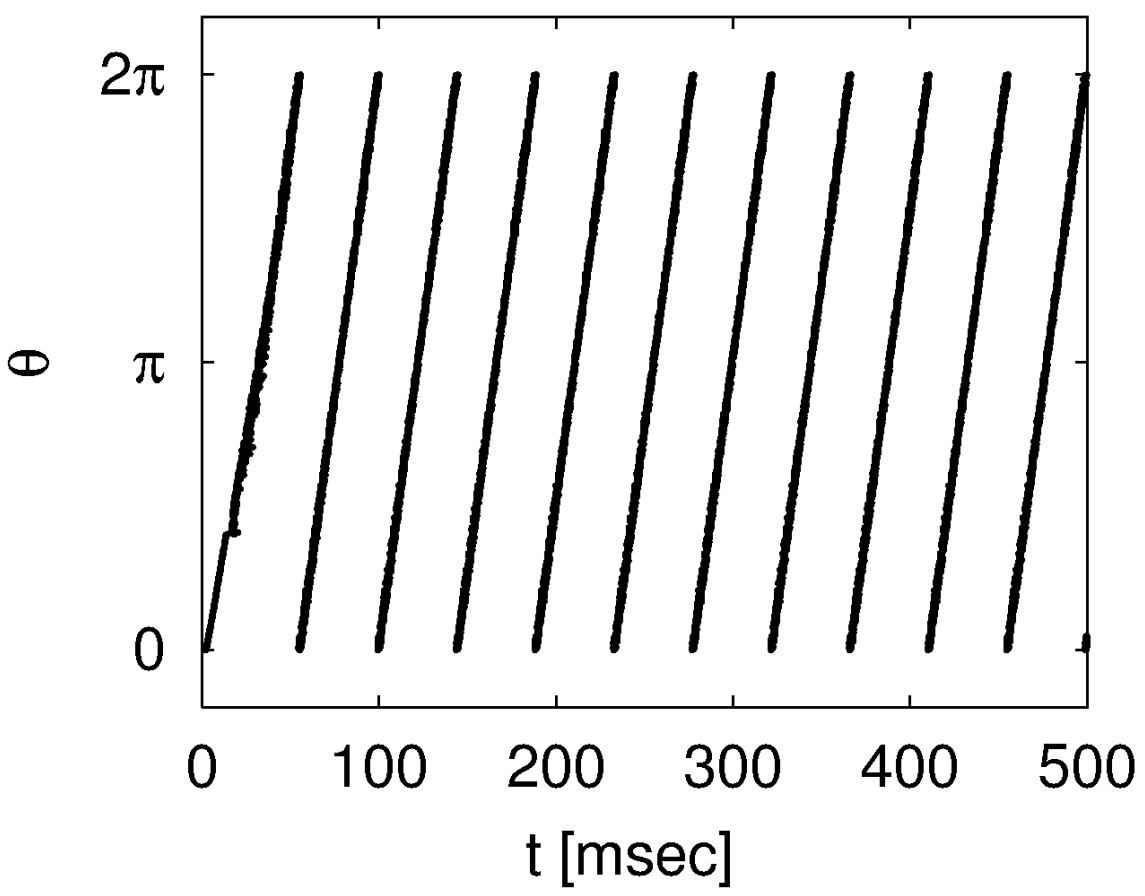

Figure 3: The result of the numerical simulation with $Q \to \infty, A_{pyr} = 20000, A_{int} = 250, P = 3$, and $N = 2000$. (**a**) The traces of firing times of neurons are plotted with points. In the interval $0 \leq t < a_{ext}T_{ext} = 12$, we apply the pulsed external electric current $I_{ext}(t)$ of the form (46) with $T_{ext} = 60$ so that the initial part of the target pattern is forced to be retrieved. (**b**) Setting the vertical axis to represent the phase variables of the target pattern $\theta_i^1$, we replot the result in (**a**).

(**a**)

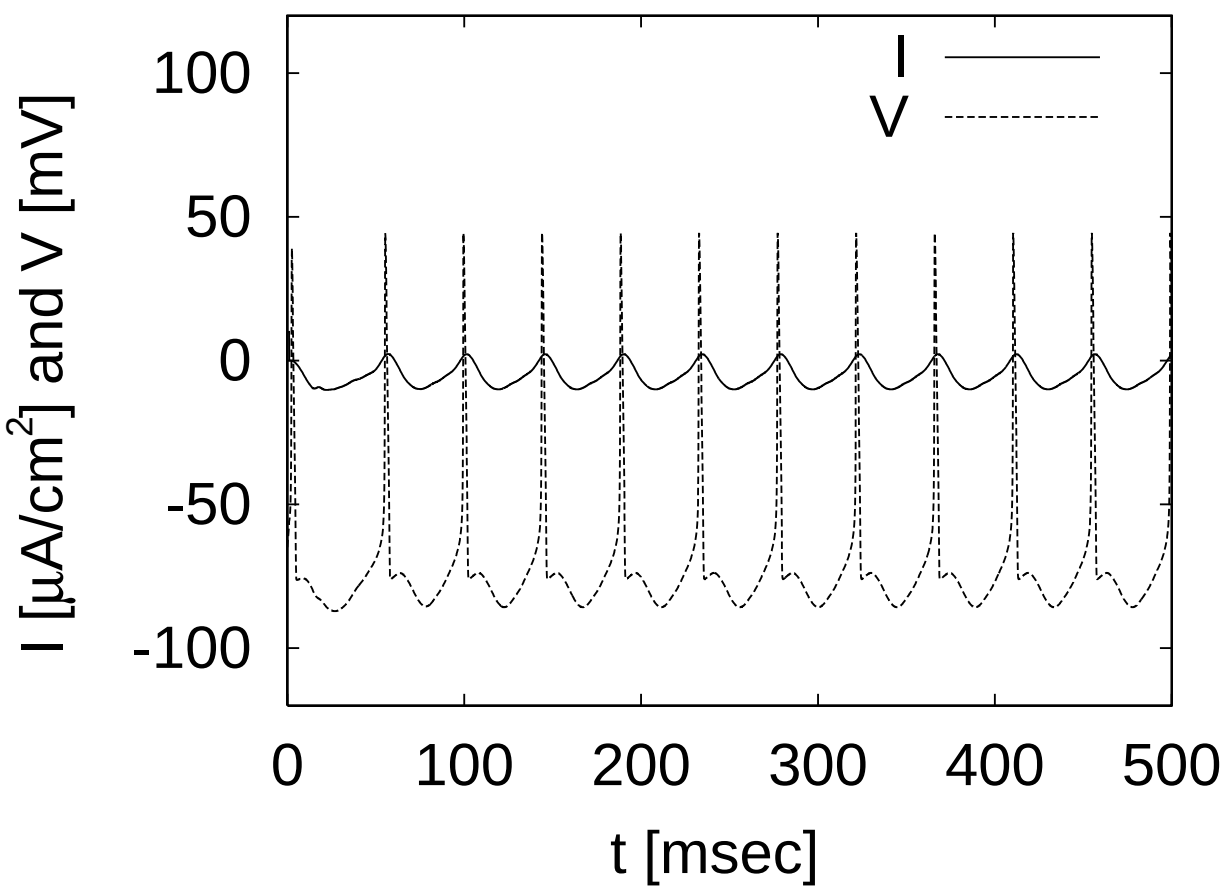

(**b**)

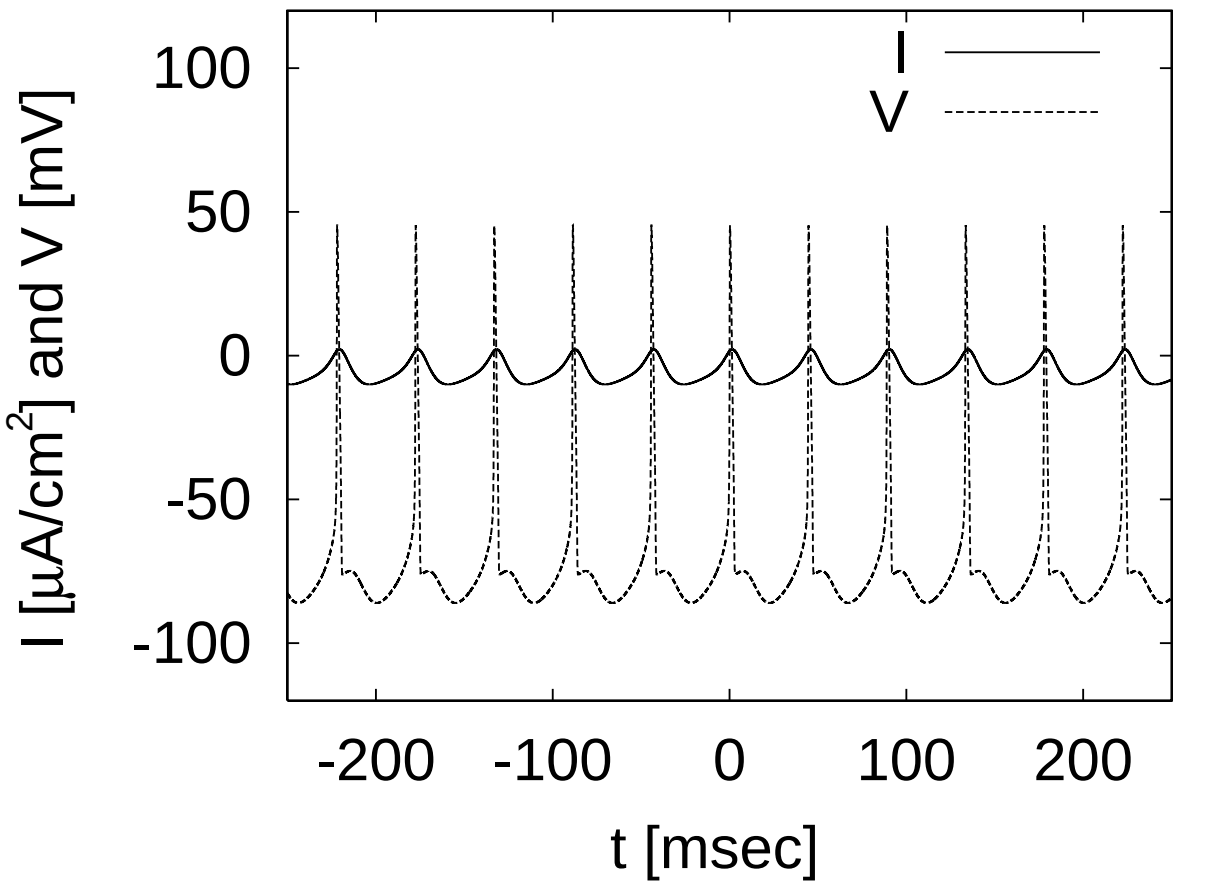

Figure 4: (**a**) Time evolution of the membrane potential $V$ of a neuron with $\theta_i^1 = 0$, which is observed in the numerical simulation in Fig. 3, is plotted together with a sum of the synaptic electric currents $I_i(t) = I_{pyr,i}(t) + I_{int}(t) + I_{ext,i}(t)$. (**b**) The result of the theoretical evaluation of the stationary behavior of a neuron. In the numerical simulation (**a**), the neuron exhibits periodical firing motion after a long time, and this periodical firing motion shows the good agreement with the result of the present analysis (**b**).

(**a**)

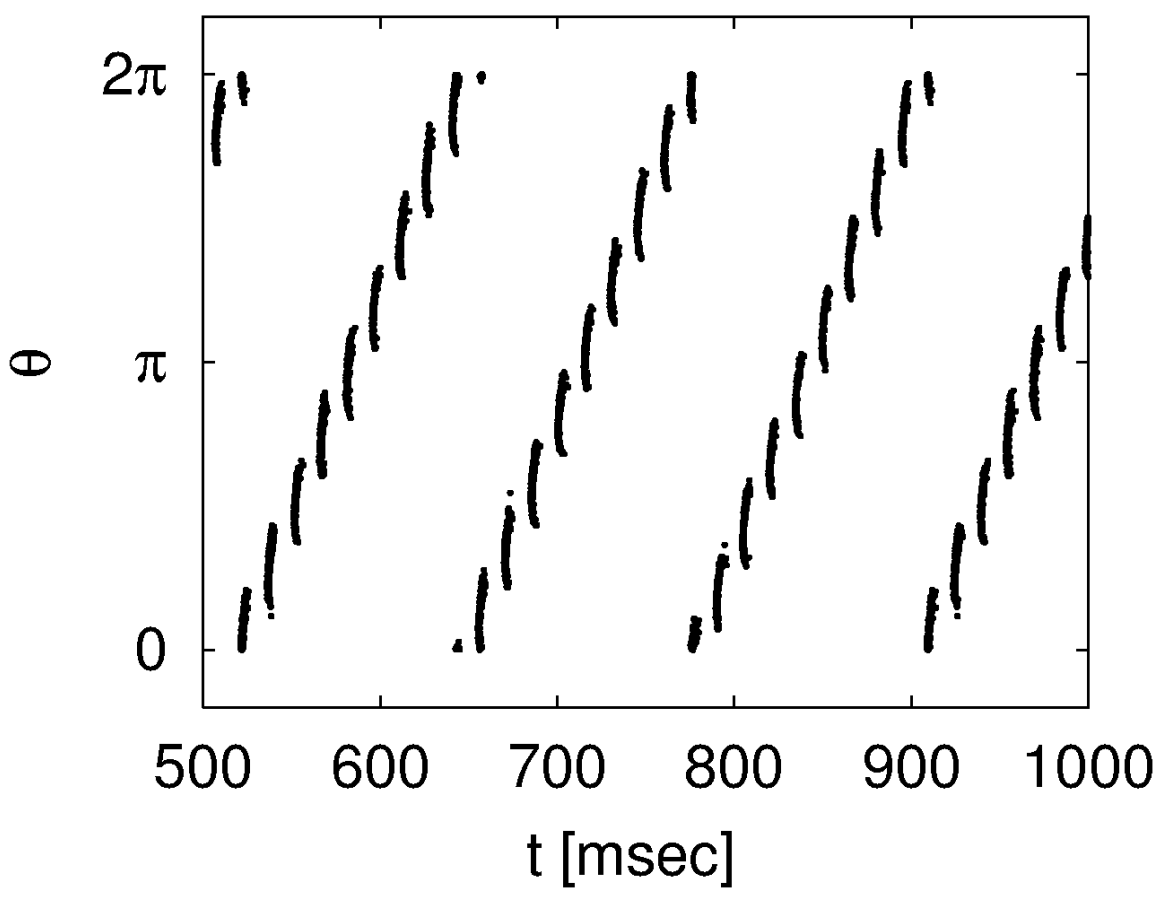

(**b**)

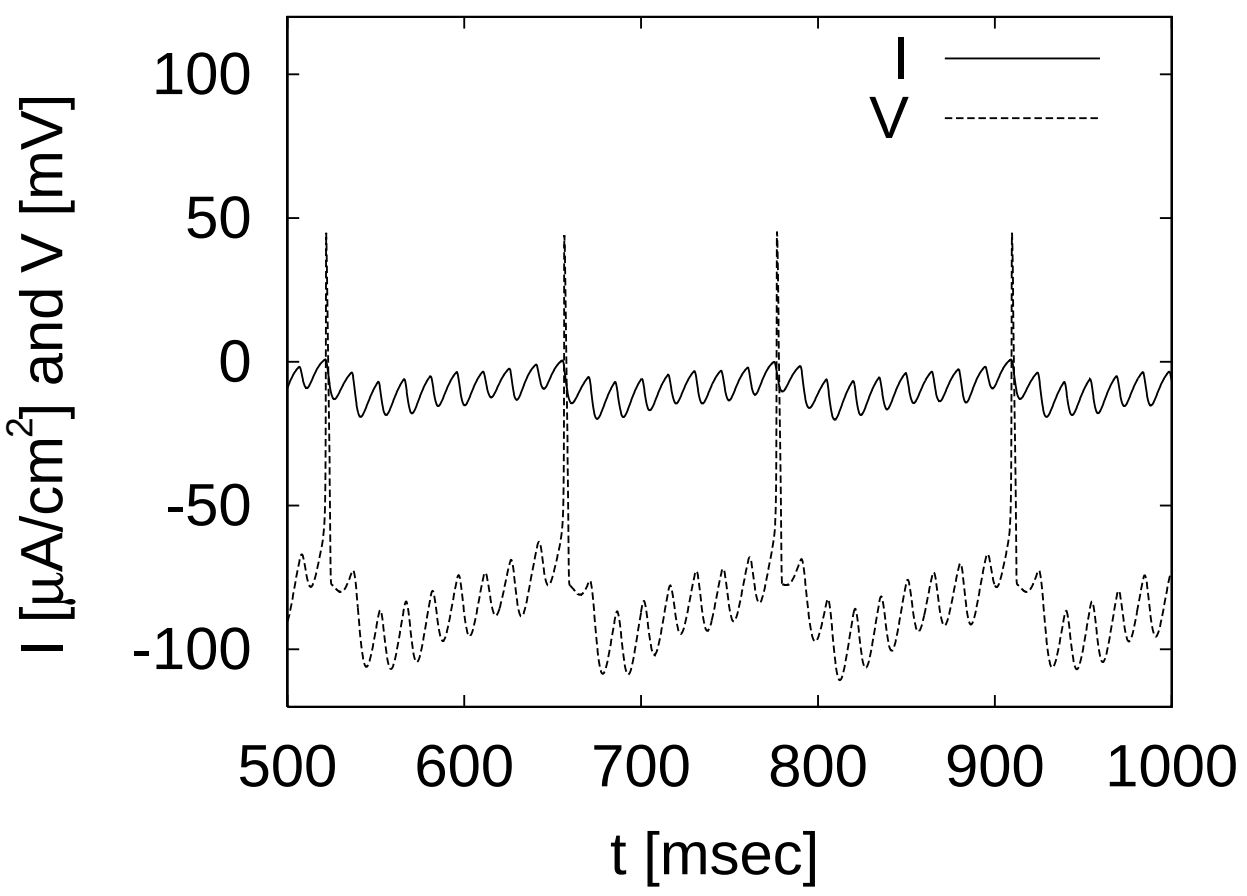

Figure 5: The result of the numerical simulation with $Q \to \infty, A_{pyr} = 20000, A_{int} = 1250, P = 3,$ and $N = 2000$. (**a**) The traces of firing times of neurons are plotted with points. Note that the vertical axis represents the phase variables of the target pattern $\theta_i^1$. (**b**) Time evolution of the membrane potential $V$ of a neuron with $\theta_i^1 = 0$ is plotted together with a sum of the synaptic electric current $I_i(t) = I_{pyr,i}(t) + I_{int}(t) + I_{ext,i}(t)$.

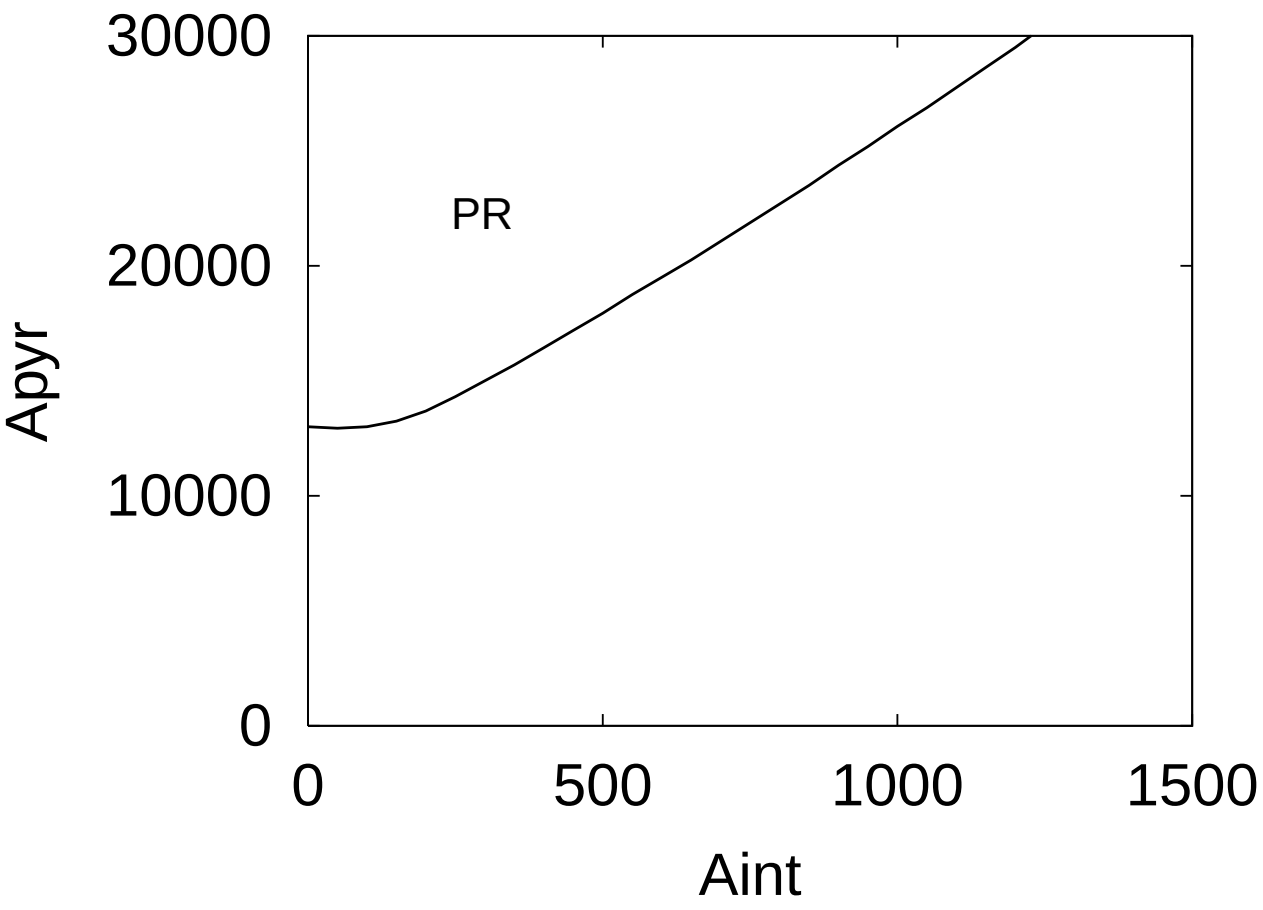

Figure 6: $A_{int} - A_{pyr}$ phase diagram showing the condition for the occurrence of the perfect retrieval in the case of $Q \to \infty$. In the region represented by PR, the perfect retrieval is realized since the period $\tilde{T}$ is successfully evaluated in the present analysis(see text). Outside the region represented by PR, the other type of stationary state is realized.

(**a**)

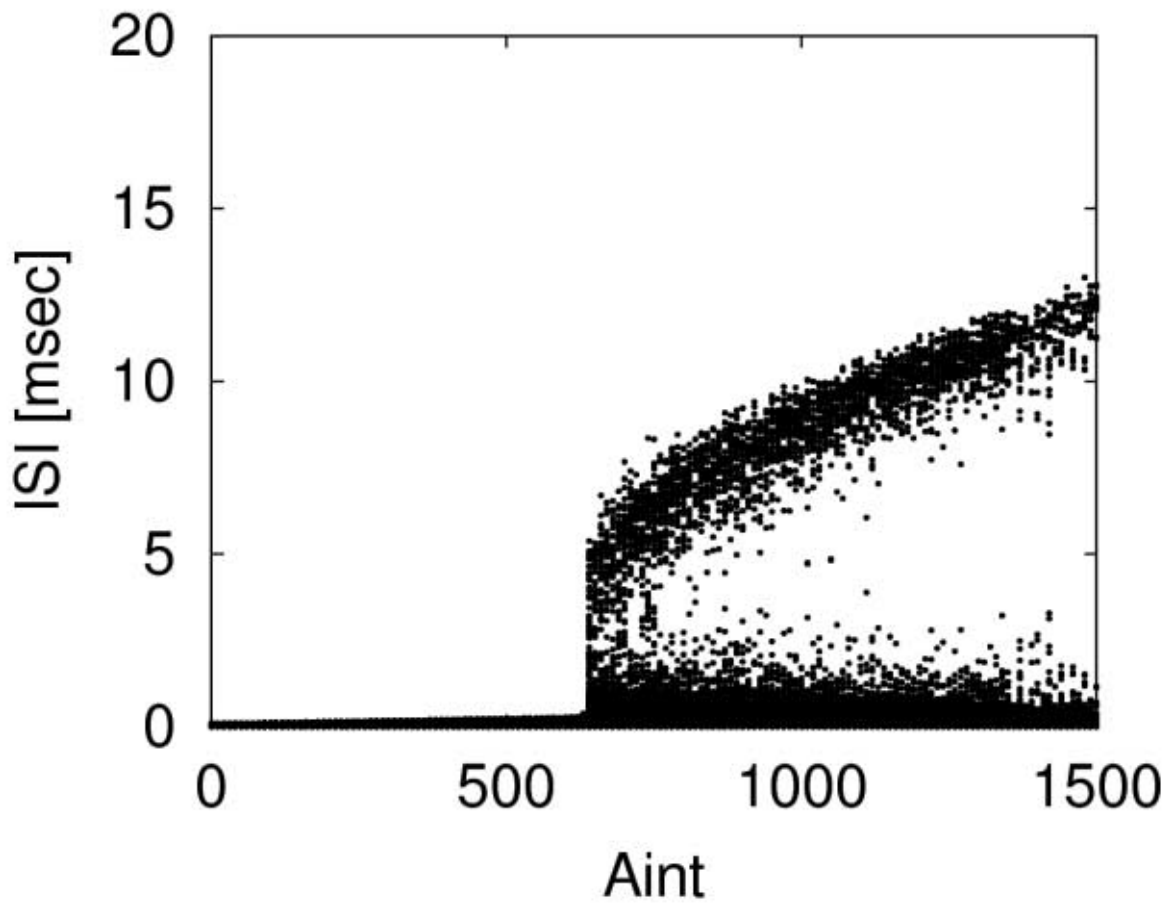

(**b**)

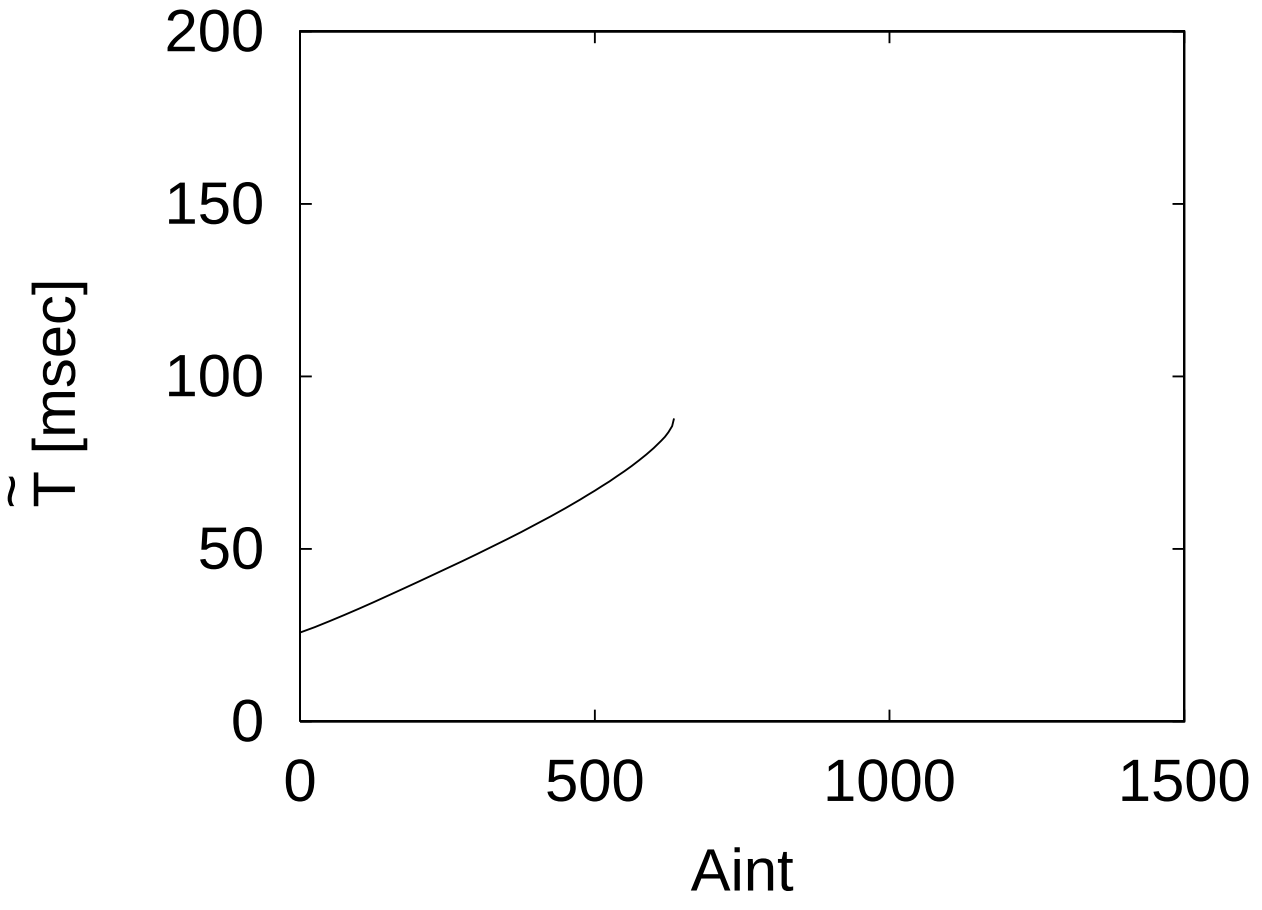

Figure 7: (**a**) A distribution of the inter spike intervals (ISIs) is plotted for the various value of $A_{int}$ in the case of $Q \to \infty, A_{pyr} = 20000, P = 1,$ and $N = 2000$. See text for the definition of the ISIs we compute. (**b**) $A_{int}$-dependence of the period $\tilde{T}$ obtained from the present analysis. We see the phase transition at the critical intensity of global inhibition $A^c_{int} \sim 630$, beyond which perfect retrieval is impossible.

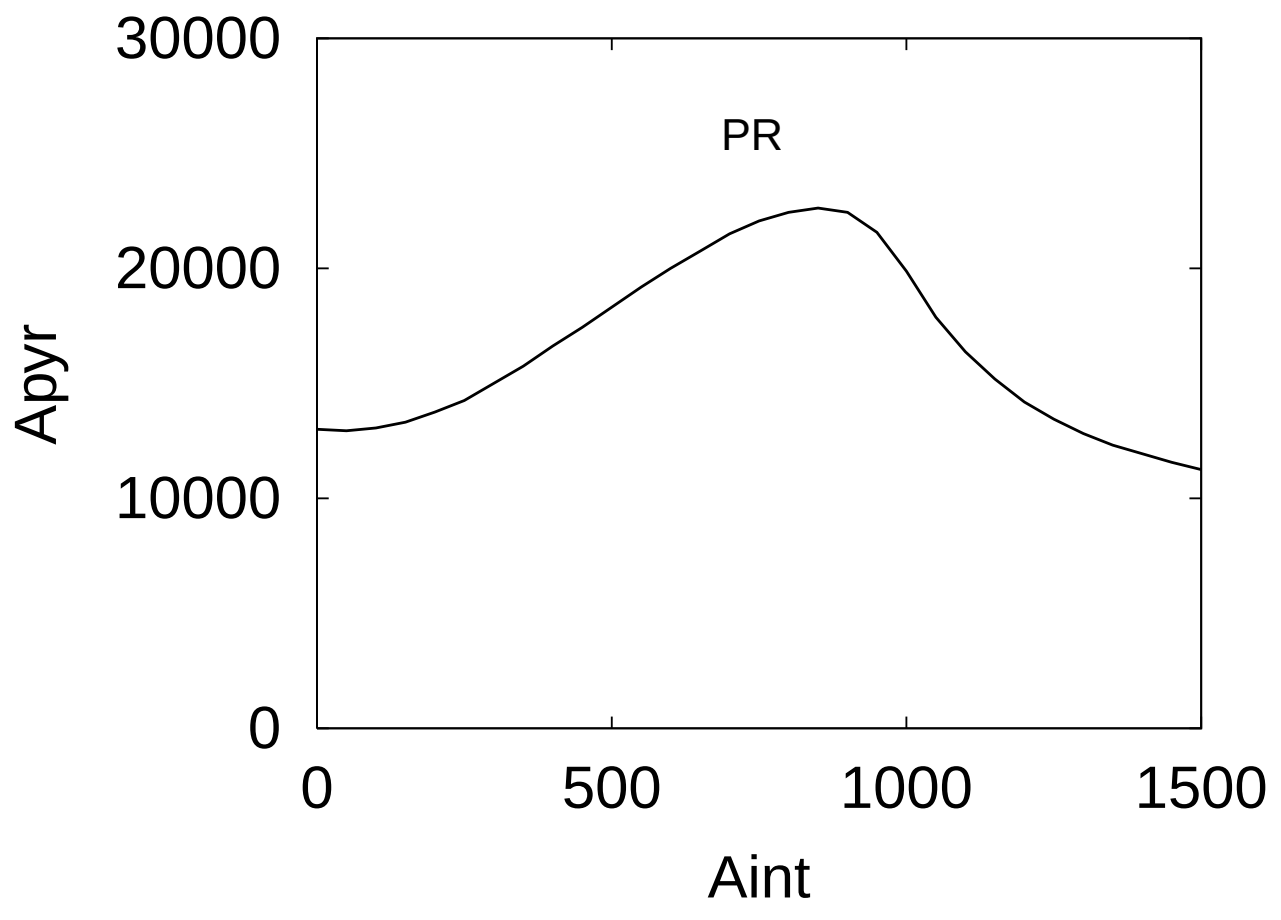

Figure 8: Same as Fig. 6, except that $Q = 10$.

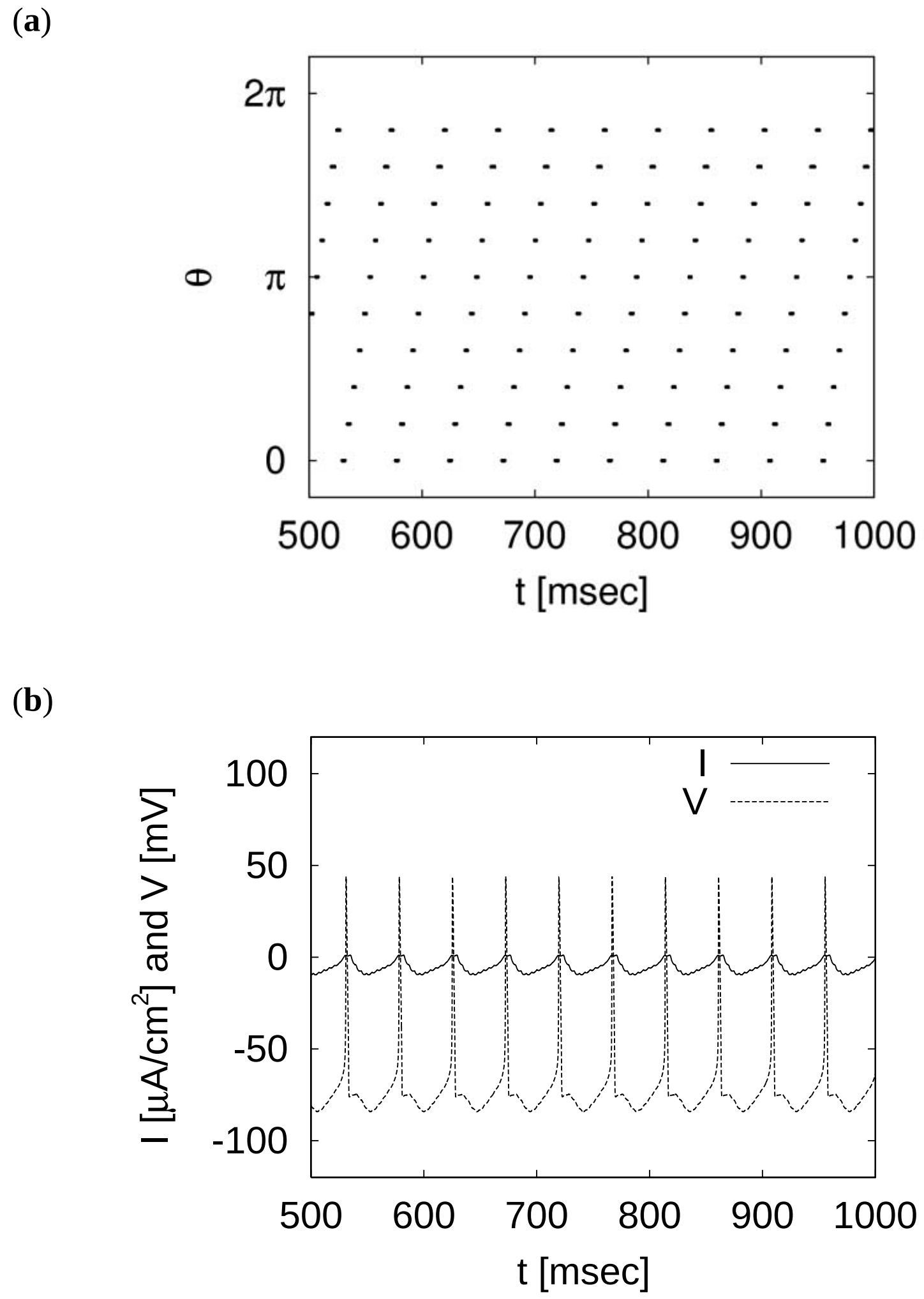

Figure 9: Same as Fig. 5, except that $Q = 10, A_{pyr} = 17000,$ and $A_{int} = 250$. Note that each point in (**a**) shows firings of $\sim N/Q$ neurons, because a cluster of neurons with the same value of $\theta_i^1$ fire synchronously.

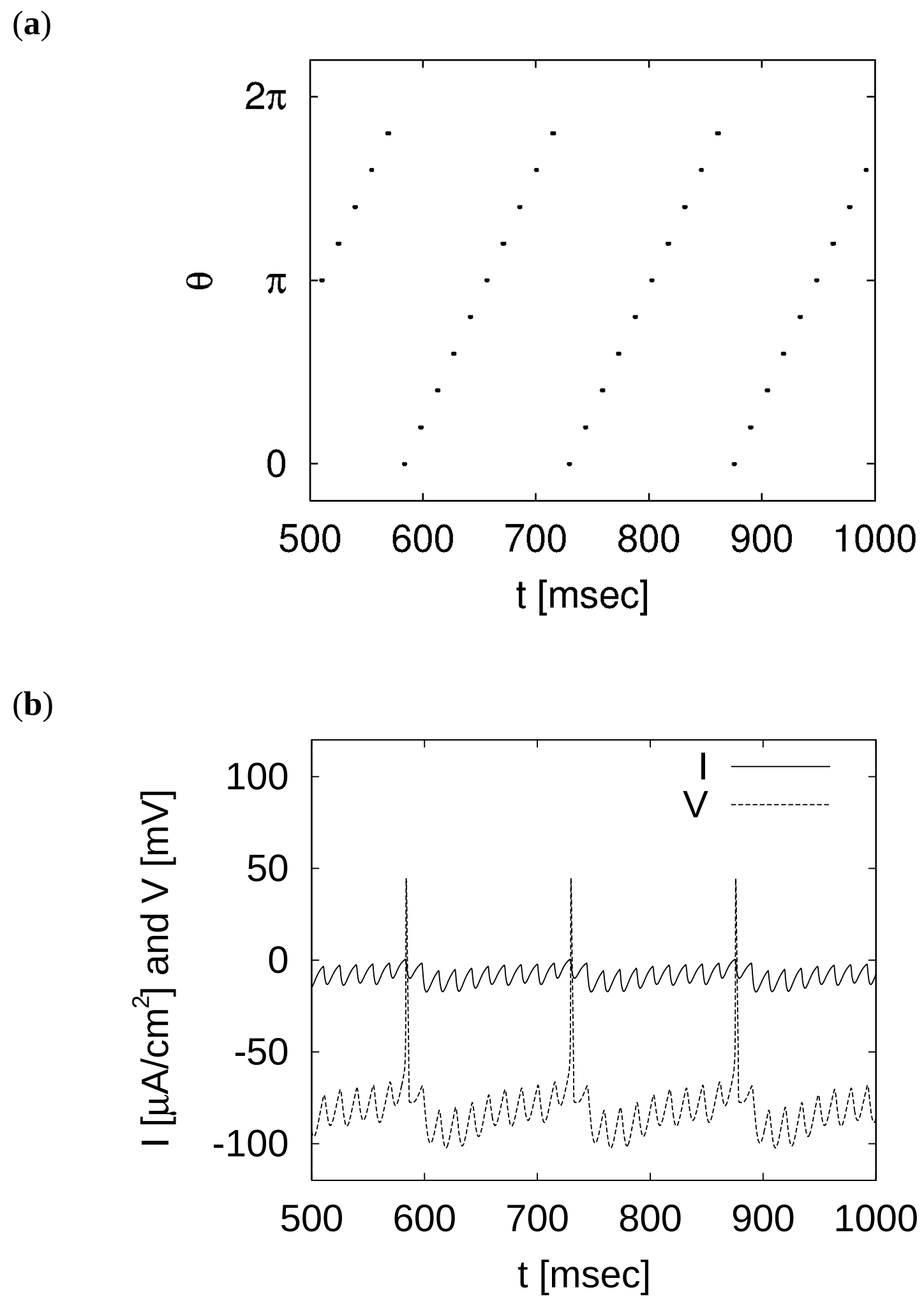

Figure 10: Same as Fig. 5, except that $Q = 10, A_{pyr} = 17000,$ and $A_{int} = 1250$.

(a)

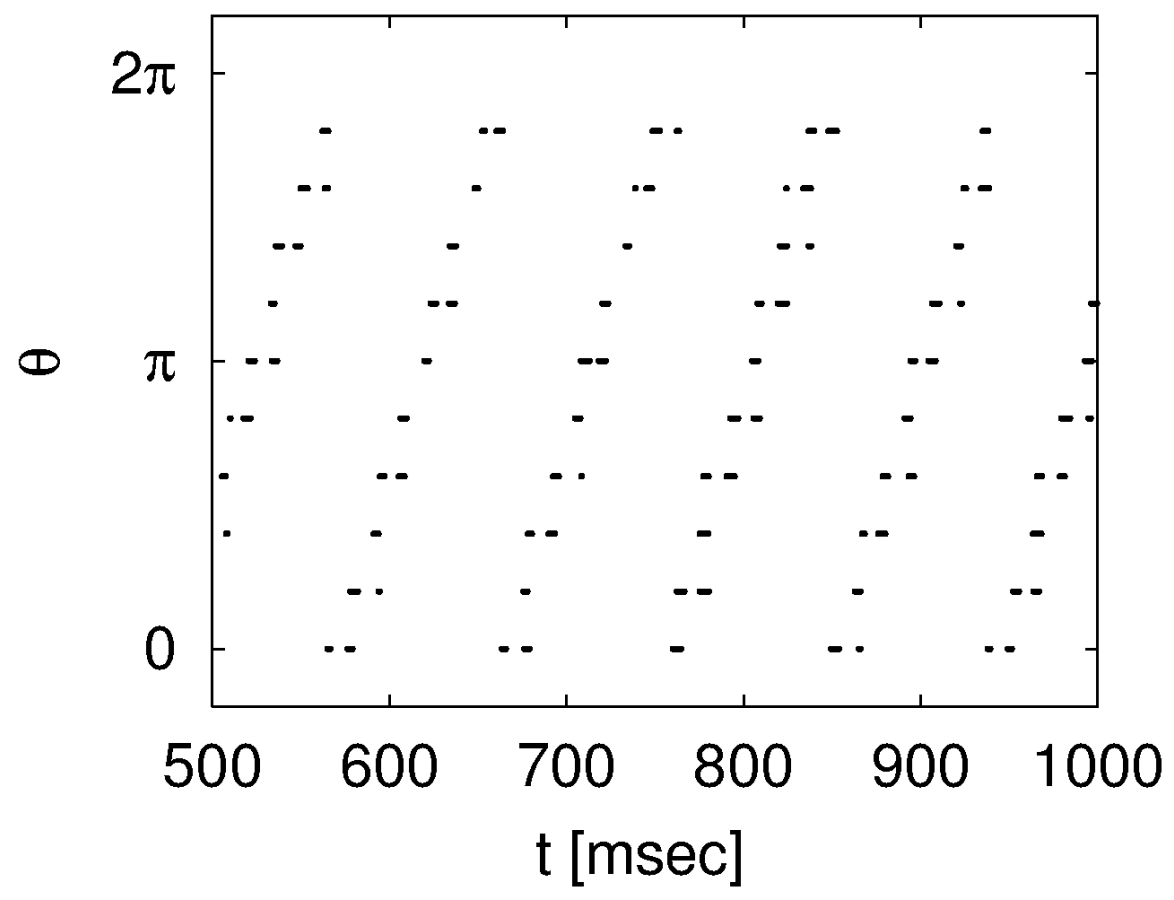

(b)

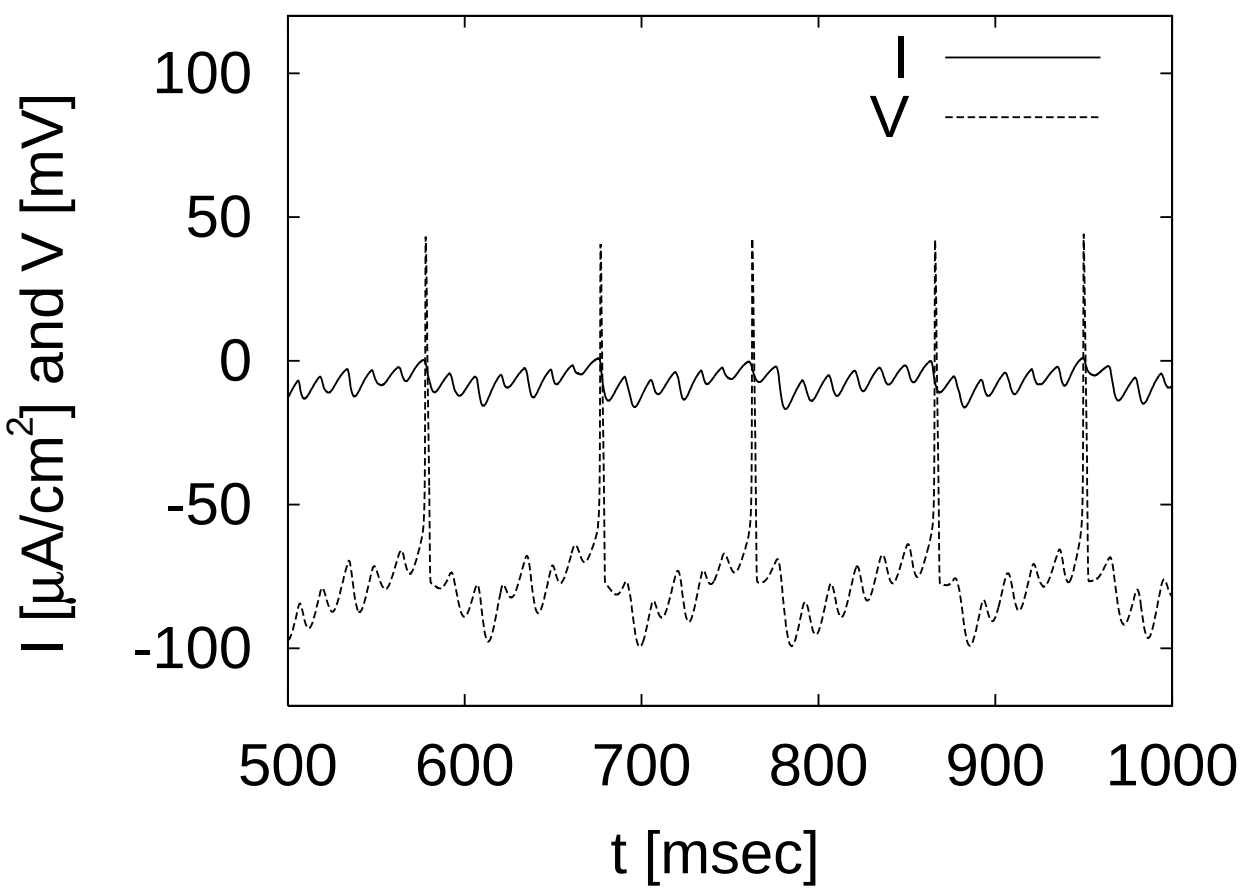

Figure 11: Same as Fig. 5, except that $Q = 10, A_{pyr} = 17000,$ and $A_{int} = 750$.

(**a**)

(**b**)

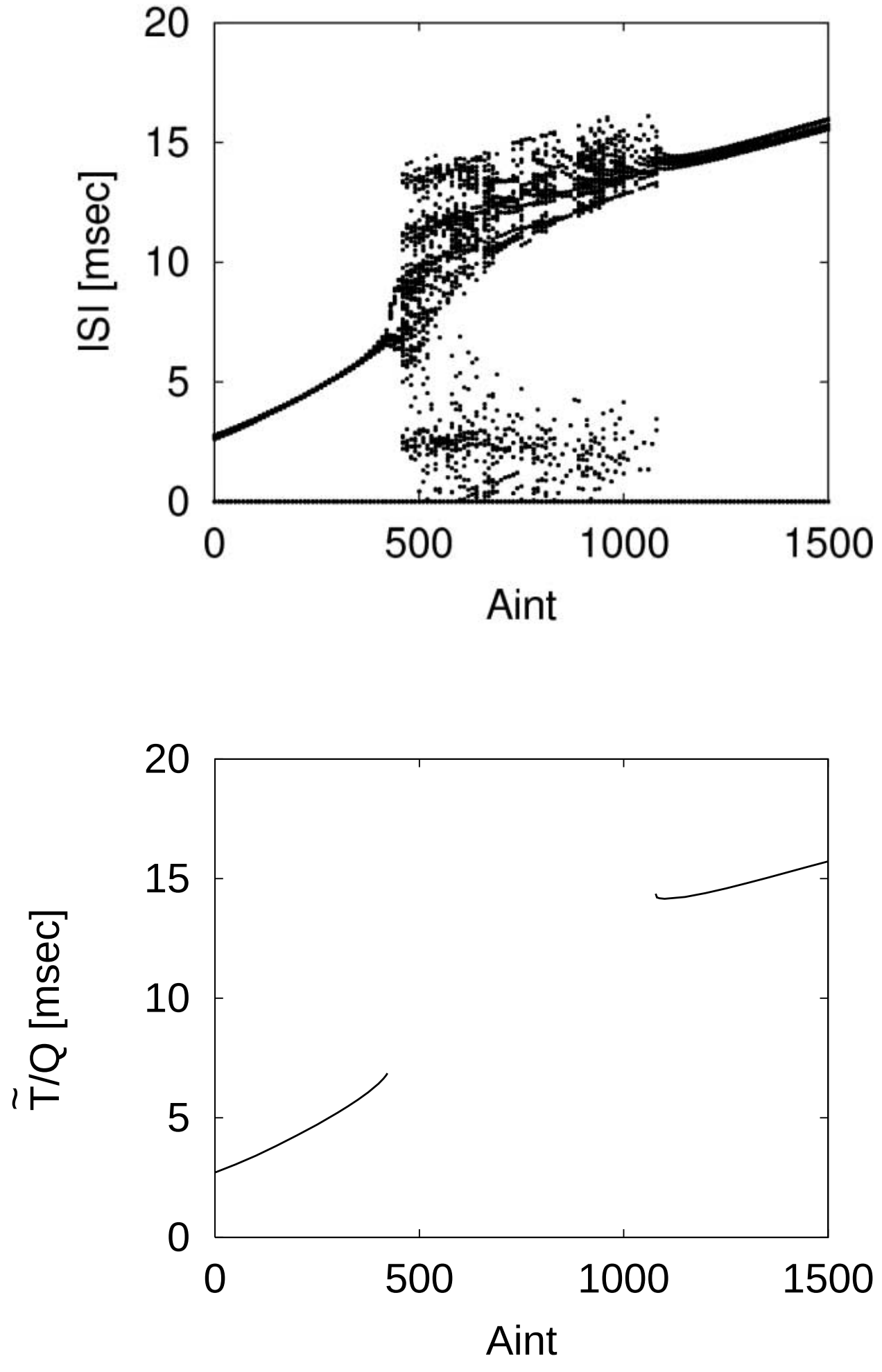

Figure 12: (**a**) Same as Fig. 7(**a**), except that $Q = 10$ and $A_{pyr} = 17000$. (**b**) $A_{int}$-dependence of $\tilde{T}/Q$ obtained from the present analysis is plotted. We see the two types of phase transitions at the critical intensity of global inhibition $A^c_{int}(1) \sim 420$ and $A^c_{int}(2) \sim 1080$. In the interval $A^c_{int}(1) \leq A_{int} \leq A^c_{int}(2)$, the perfect retrieval is impossible.

(**a**)

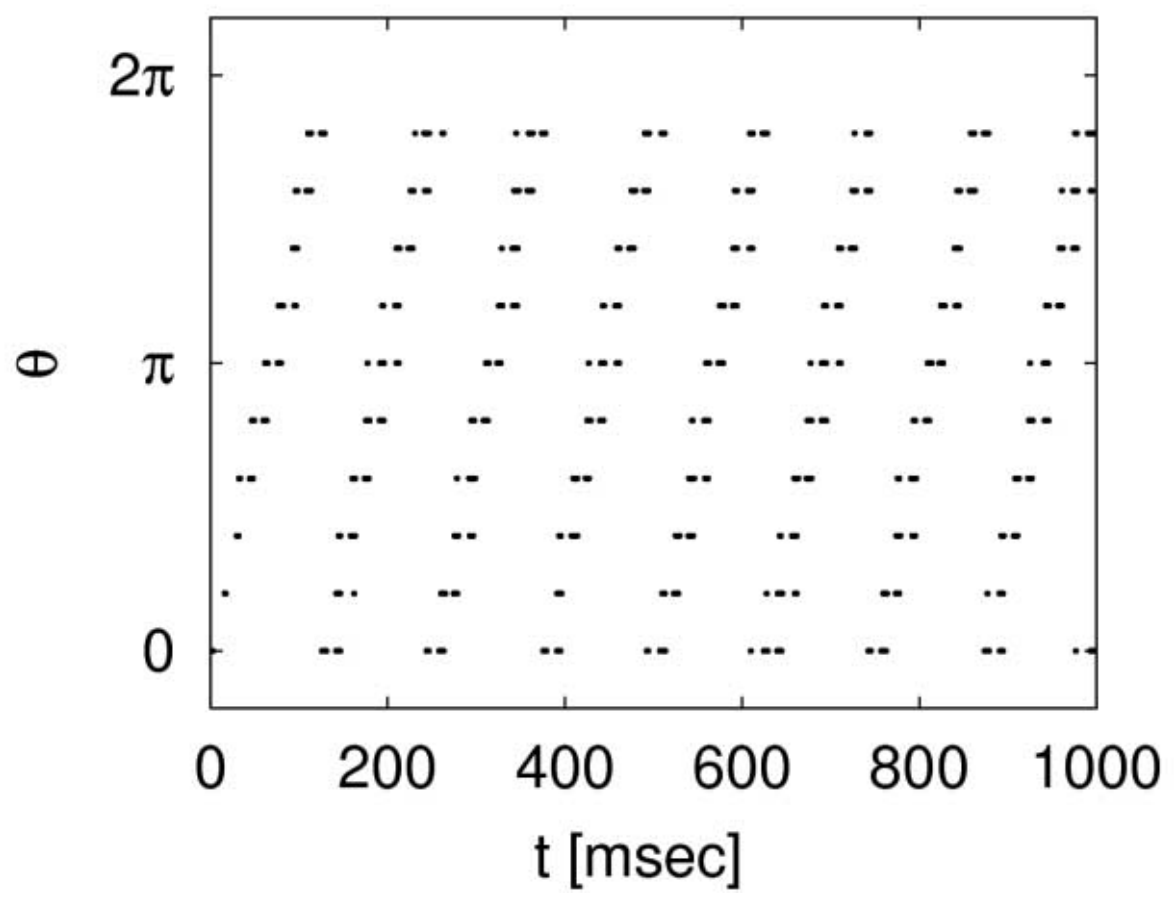

(**b**)

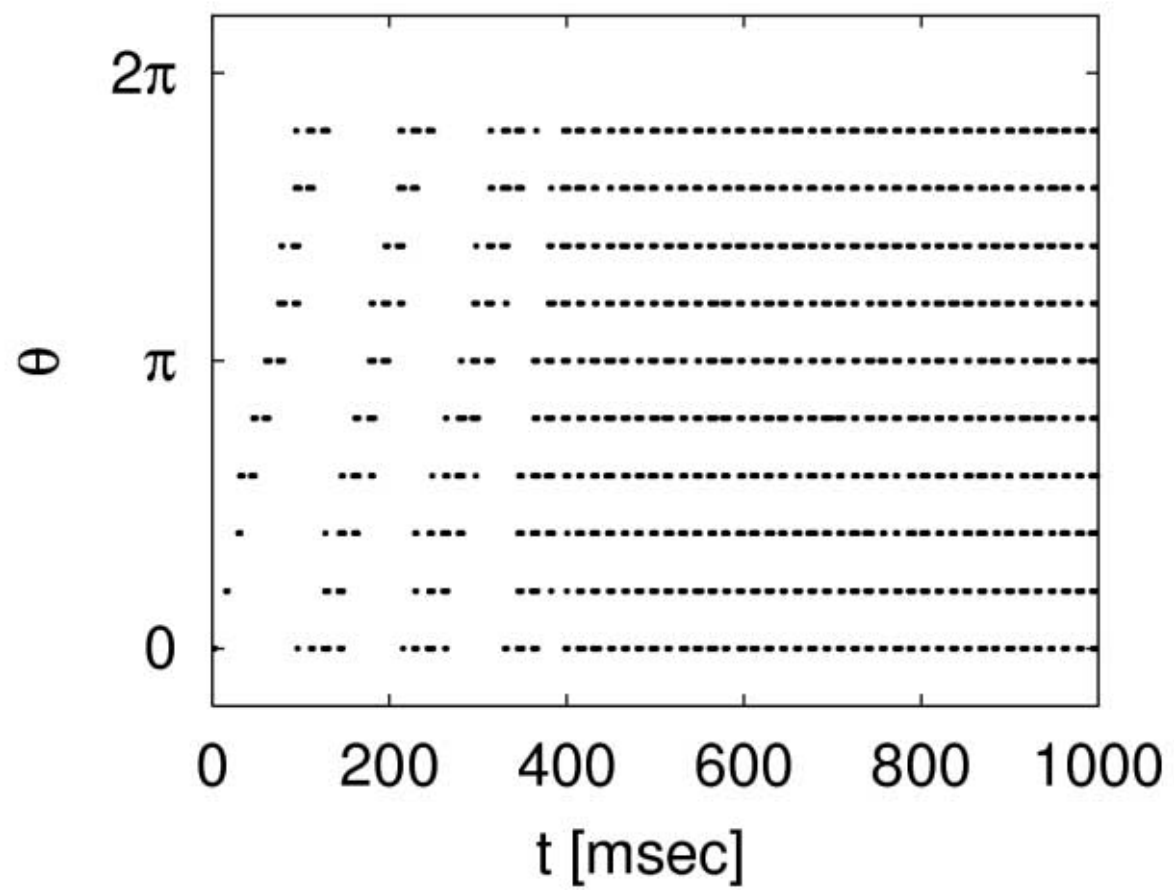

Figure 13: We estimate the storage capacity $\alpha^c = P^c/N$ based on the numerical simualtions with $N = 2000$. As a number of stored patterns $P$ increases, the distribution of the firing times of neurons becomes wider as a result of the increase in the size of the crosstalk term. In the case of (**a**), pattern retrieval is still successful since the loading rate $\alpha = P/N = 0.007$ is less than the storage capacity $\alpha^c$. On the other hand, in the case of (**b**), pattern retrieval is impossible since the loading rate $\alpha = 0.01$ is beyond the storage capacity $\alpha^c$. The value of parameters are $Q = 10, A_{pyr} = 17000,$ and $A_{int} = 1250$.